\documentclass[letterpaper, 10 pt, journal, twoside]{IEEEtran}
\usepackage[colorlinks=true,bookmarksopen,bookmarksnumbered, linkcolor=black, citecolor=black,urlcolor=blue]{hyperref}
\usepackage{color}
\usepackage{stfloats}
\usepackage{graphicx} 
\usepackage{subcaption}
\usepackage{amsmath} 
\usepackage{amssymb}  
\usepackage{csquotes}

\usepackage{array}
\usepackage{url} 
\usepackage{enumerate}
\usepackage{multirow} 
\usepackage{float}
\usepackage{paralist}
\usepackage{amsfonts}
\usepackage{longtable}

\newcolumntype{L}[1]{>{\raggedright\let\newline\\\arraybackslash\hspace{0pt}}m{#1}}
\newcolumntype{C}[1]{>{\centering\let\newline\\\arraybackslash\hspace{0pt}}m{#1}}
\newcolumntype{R}[1]{>{\raggedleft\let\newline\\\arraybackslash\hspace{0pt}}m{#1}}

\DeclareMathOperator*{\argmin}{arg\,min}

\graphicspath{ {images/} }

\newcommand{\nostarnote}[1]{}
\newcommand{\baad}[1]{} 
\usepackage[size=small]{caption}
\newcommand{\ie}{\textit{i.e.}}
\newcommand{\eg}{\textit{e.g.}}
\newcommand{\etal}{\textit{et al.}} 

\usepackage{todonotes}
\usepackage{enumitem}
\usepackage{makecell}

\ifCLASSINFOpdf
\else
\fi
\title{\LARGE \bf
LightViz: Autonomous Light-field Surveying and Mapping for Distributed Light Pollution Monitoring}

\begin{document}

%
%
%

\author{Sheng-En Huang$^\ast$, Kazi Farha Farzana Suhi$^\diamond$, and Md Jahidul Islam$^\ast$ \\
{
\small $^\ast$RoboPI Laboratory, Department of ECE, University of Florida, FL 32611, US \\
\vspace{-1mm}
$^\diamond$College of Design Architecture Art and Planning, University of Cincinnati, OH 45221, US
} \\
\vspace{1mm}
{
\small 
Project page: \url{https://robopi.ece.ufl.edu/lightviz.html}\\
Email: \tt huang.sh@ufl.edu, jahid@ece.ufl.edu, suhikf@mail.uc.edu
\vspace{-7mm}
}
\thanks{This pre-print is accepted for publication at the Springer Nature Environmental Monitoring and Assessment (EMA); Link: \url{https://doi.org/10.1007/s10661-025-13862-5}.} 
}

\maketitle


\begin{abstract}
Existing technologies for distributed light-field mapping and light pollution monitoring (\textbf{LPM}) rely on either remote satellite imagery or manual light surveying with single-point sensors such as SQMs (sky quality meters). These modalities offer low-resolution data that are not informative for dense light-field mapping, pollutant factor identification, or sustainable policy implementation. In this work, we propose \textbf{LightViz} -- an interactive software interface to survey, simulate, and visualize light pollution maps in real-time. As opposed to manual error-prone methods, LightViz (\textbf{i}) automates the light-field data collection and mapping processes; (\textbf{ii}) provides a platform to simulate various light sources and intensity attenuation models; and (\textbf{iii}) facilitates effective policy identification for conservation. To validate the end-to-end computational pipeline, we design a distributed light-field sensor suite, collect data on Florida coasts, and visualize the distributed light-field maps. In particular, we perform a case study at St. Johns County in Florida, which has a two-decade conservation program for lighting ordinances. The experimental results demonstrate that LightViz can offer high-resolution light-field mapping and provide interactive features to simulate and formulate community policies for light pollution mitigation. We also propose a mathematical formulation for `light footprint' evaluation, which we integrated into LightViz for targeted LPM in vulnerable communities. A test-case of the LightViz software release is available at: \url{https://github.com/uf-robopi/LightViz}.
\end{abstract}


%
\IEEEpeerreviewmaketitle

%
%
%
%

\section{Introduction}
\IEEEPARstart{O}{ver} $80\%$ of the world and almost all U.S. cities and urban skylines are affected by artificial light pollution. With the rapid growth of metropolitan areas and coastal cities, alarming levels of light pollution result in significant long-term consequences for both humans and other animals~\cite{rodrigo2023light}. Contemporary research over the past decade has shown that light pollution negatively impacts human health by disrupting natural sleep cycles, causing chronic sleep deprivation, which in turn increases risks for high blood pressure, exhaustion, and depression~\cite{BENFIELD201867,gaston2013ecological}. Moreover, various animals such as sea turtles, bees, squirrels, birds, and insects experience the disruption of  their nocturnal patterns ~\cite{raap2017disruptive,navara2007dark} by long periods of light exposure, direct glare, and sky glow. Studies have evidenced that these severely affect their natural reproductive cycles, cause hormonal dysfunctions, and trigger serious long-term health issues~\cite{boyes2021light,grubisic2019light,owens2020light}. 

For instance, light pollution in beachfront areas has caused significant habitat loss for sea turtles across the globe~\cite{lohmann2017orientation,truscott2017effect}. Female nesting turtles tend to avoid brightly lit beaches and the positive phototaxis of hatchlings—which is essential for their correct trajectory from nest to sea—is interfered with by artificial lighting inland. The results of these impacts are so severe that the Endangered Species Act~\cite{valdivia2019marine} has listed all six sea turtles as \emph{endangered} in the U.S. coastal waters. To mitigate the adverse consequences of light pollution on human health and wildlife habitats, long-term community initiatives and sustainable policy decisions are essential. The development of next-generation tools for measuring and monitoring light pollution at the county level is needed in order to address this crisis. Furthermore, encouraging community participation and awareness through interactive system interfaces for crowd-sourcing \textit{light footprint} measurements will facilitate effective policy implementation for conservation.

 \begin{figure}[t]
\centering
\begin{subfigure}{0.5\textwidth}
    \centering
        \includegraphics[width=\linewidth]{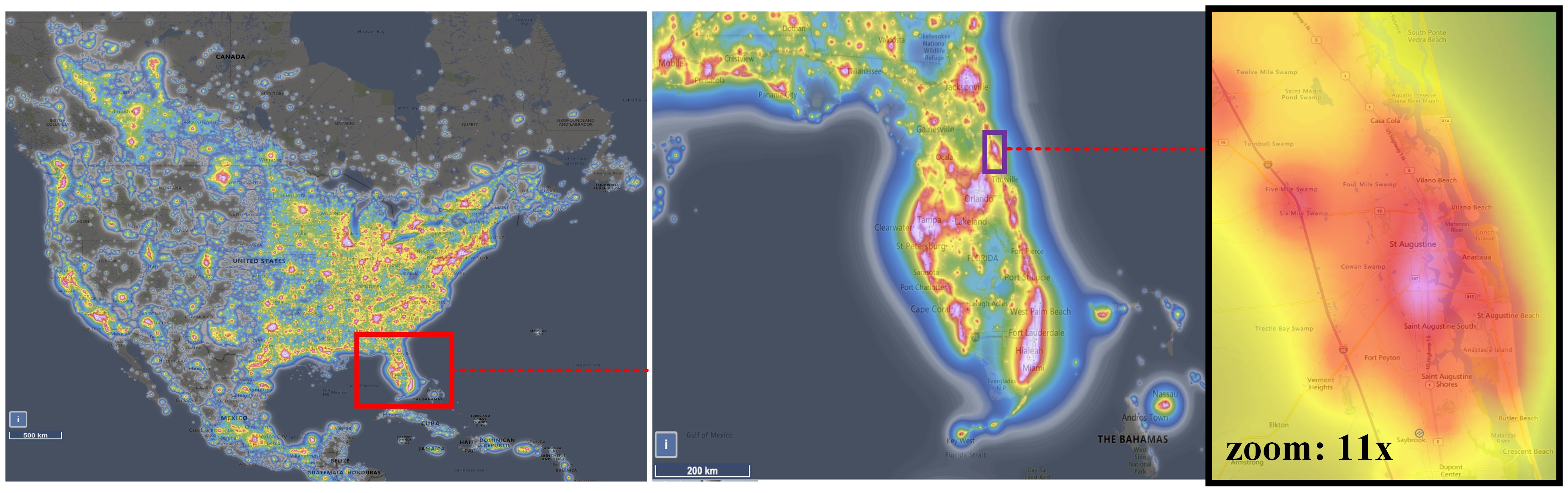}%
        \vspace{-2mm}
        \caption{Low-resolution light pollution maps from existing interfaces: VIIRS and World Atlas data~\cite{falchi2016new,nurbandi2016using} are shown for St. Augustine, FL, US.}
        \label{fig:intro_a}
    \end{subfigure}

    \begin{subfigure}{0.5\textwidth}
    \centering
    \vspace{1mm}
    \includegraphics[width=\linewidth]{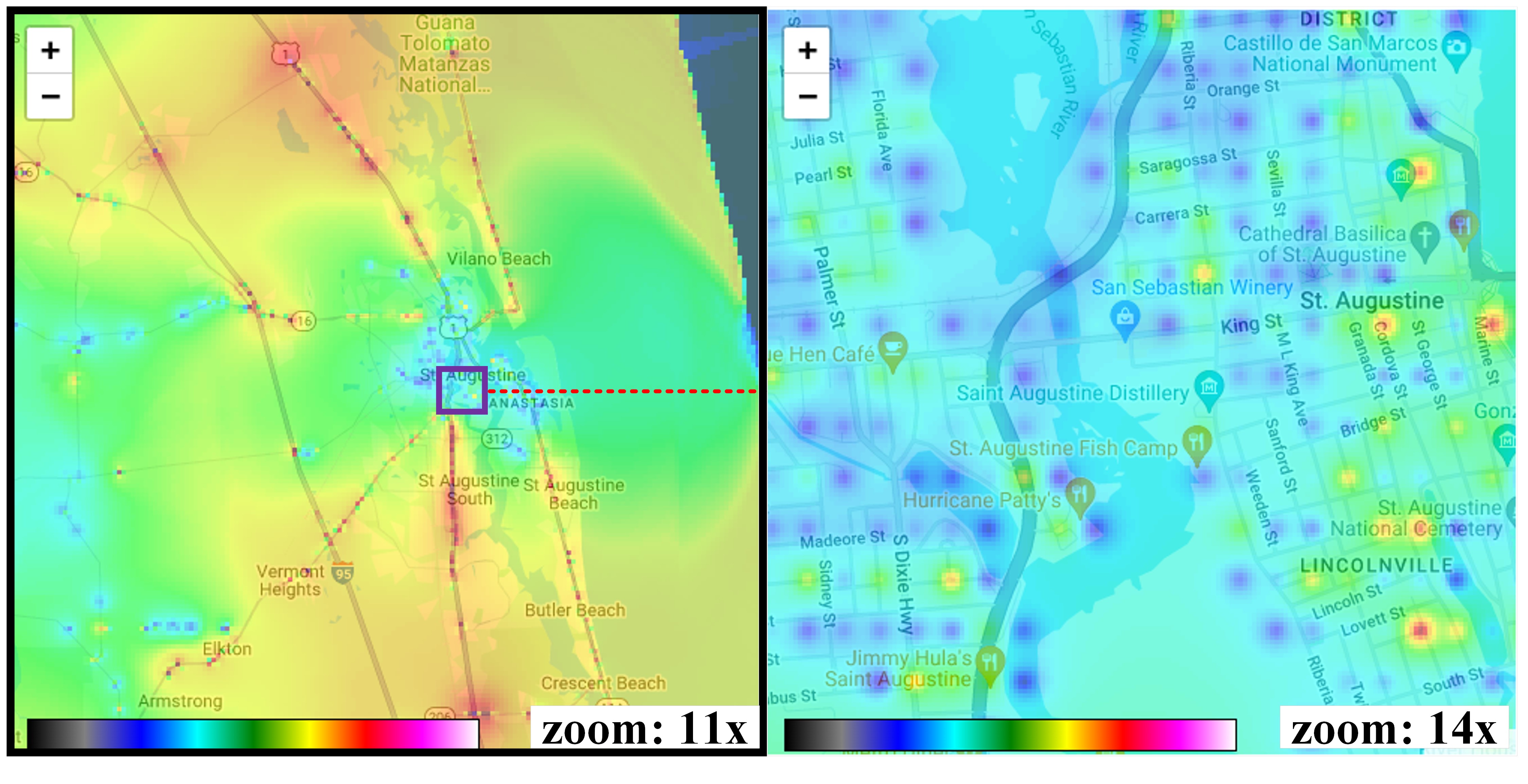}%
        \vspace{-3mm}
        \caption{High-resolution map generated by LightViz embed fine-grained information: $3\times$ higher zoom level is shown for the same region.}
        \label{fig:intro_b}
    \end{subfigure} 
	\vspace{-4mm}
	\caption{Effectiveness of the \textbf{LightViz} interface is demonstrated by distributed light-field mapping for LPM (light pollution monitoring).}
	\label{fig:intro}
 \vspace{-4mm}
\end{figure}

Standard ambient light sensors (TSL2591~\cite{karpinska2022device}, SQM~\cite{mander2023measure}) capture single-point radiance measurements,  while the satellite imagery (NASA VIIRS~\cite{falchi2016new}, World Atlas~\cite{nurbandi2016using}) generate continental-scale maps that only identify pollution-prone areas; see Fig.~\ref{fig:intro_a}. However, these methods fail to identify pollutant sources and quantify their  individual and collective `{light footprint}'. Thus, measuring long-term impacts on humans or other habitats remains challenging. Contemporary works~\cite{windle2018robotic,ludvigsen2018use,massetti2022monitoring} use robotics and automated tools for high-resolution light surveying with promising results. However, these efforts are not scalable as an LPM tool for effective policy implementation at a county, city, or state level. 

In recent years, UAVs equipped with simple RGB cameras have gained prevalence for monitoring urban lighting installations and rendering detailed maps~\cite{zhang2024evaluation,bouroussis2020assessment}. Targeted aerial data and geospatial maps can help identify light pollution hotspots and quantify upward light emissions accurately. In particular, robust frameworks have been developed for night-ground brightness (NGB) monitoring~\cite{massetti2022monitoring} by combining RGB image indices with ground-based SQM measurements. However, UAV flight times are limited, and night-time image data registration with single-point on-ground sensors requires meticulous calibration~\cite{fiorentin2019calibration}. The considerable logistics and time needed for high-resolution measurements further constrain comprehensive LPM at scale~\cite{pHodor2022detecting}. Thus, the image data remains sparse for large-scale deployments; that is, although the generated map is high-resolution, the coverage area per pixel is still high. Such maps are not interactive to investigate mitigation policies in a given community, and are not repeatable for simulating lighting strategies for LPM research.

We address these long-held challenges of distributed LPM by introducing \textbf{LightViz} -- an interactive interface designed to survey, simulate, and visualize light-fields and light pollution maps in real-time, see Fig.~\ref{fig:intro_b}. The core strength of LightViz lies in its integration of light source placement and a light attenuation model, enabling the generation of light-field maps with fine-grained detail and local continuity. Existing light-field measurements based on satellite imagery~\cite{falchi2016new,nurbandi2016using} or ground-based single-point sensor nodes~\cite{karpinska2022device, mander2023measure} suffer from low sampling rates, varying resolution, and high latency. The considerable cost and time required for high-resolution measurements further constrain comprehensive LPM at scale. LightViz overcomes these limitations by: (\textbf{i}) offering end-to-end light-field rendering; (\textbf{ii}) simulating light sources to produce dense maps at significantly higher resolutions; (\textbf{iii}) capturing fine-grained variations in local light-field data; and (\textbf{iv}) addressing latency issues in global map generation.

Existing light pollution interfaces~\cite{falchi2016new,nurbandi2016using} render averaged color gradients, which provide poor spatial resolution when zoomed in. In contrast, LightViz produces high-resolution maps with street-level information, as shown in Fig.~\ref{fig:intro}. Another important feature of LightViz is the adjustable placement of light sources on an interactive basemap, which facilitates the formulation of lighting ordinances for policy-making. Specifically, LightViz identifies hotspots of light pollution and vulnerable areas using a gradient heatmap to analyze various lighting configurations. Thus, it can potentially assist in revising street lighting regulations, including controlling lighting duration, optimizing street light placement, and promoting energy-efficient lighting fixtures. These features are absent in state-of-the-art (SOTA) software interfaces~\cite{elsahragty2015assessment} that use various GIS models to formulate lighting strategies for LPM.

For performance validation, we first design and develop a novel light-field sensing module that facilitates long-term radiance sensing (standalone mode) and enables dense light-field sensing in a given area (mobile mode). Compared to traditional single-point measurements with manual surveys, these maps are more accurate, dense, and reliable -- enabling long-term LPM~\cite{huang2024darkmeter}. We deploy our sensor suit on a beachfront community in Florida and demonstrate the challenges involved in generating high-resolution distributed light-field maps at scale. These field trials validate the need for LightViz as a user interface to simulate and visualize light-fields for LPM. It is important to make the distinction that \emph{light-field maps} in the computer graphics literature is different~\cite{chen2002light} -- which focuses on the physics of light transport -- such as scattering and absorption characteristics within a scene to project ray tracing densities when light propagates through a medium. In the context of light pollution, light-field maps visualize the spatial distribution of artificial light emissions on a geographical scale.

We demonstrate the effectiveness of LightViz by conducting a case study in St. Johns County, Florida, which has a two-decade-long conservation program~\cite{lightmanagement} for mitigating light pollution along the coast. We configure over $6,000$ streetlights with individual attenuation properties and distribute those across various road types across the county. 
We find that the light-field map generated by LightViz highlights vulnerable areas, identifies pollutant light sources, and provides an efficient way to track their \textbf{light footprints} over time. Finally, we present various mathematical frameworks on LightViz for community policy identification and prospective solutions for light pollution mitigation.

\vspace{1mm}
\noindent
\textbf{Proposed hypothesis and contributions.} The primary hypothesis of this paper is that a lightweight and interactive light-field sensing and mapping system like LightViz can significantly enhance the accuracy, resolution, and scalability of LPM compared to traditional methods. By automating data collection, integrating advanced simulation and attenuation models, and providing real-time high-resolution visualizations in LightViz -- we can effectively identify pollution sources, assess their impacts on vulnerable communities, and facilitate the development of informed and sustainable light pollution mitigation policies.

As shown in Fig.~\ref{fig:intro}, LightViz contributes to the LPM technology advancement by: (\textbf{i}) offering end-to-end light-field rendering; (\textbf{ii}) simulating light sources to produce dense maps at significantly higher resolutions; (\textbf{iii}) capturing fine-grained variations in local light-field data; and (\textbf{iv}) addressing latency issues in global map generation. In addition, it facilitates exploring effective light pollution mitigation strategies. users can load existing data and simulate new data to investigate key questions such as: (\textbf{i}) How do the intensity and spatial distributions of light pollution vary across urban, suburban, and rural environments?; (\textbf{ii}) What are the primary factors contributing to light pollution in a given vulnerable area? (\textbf{iii}) How can systematic and effective mitigation strategies be developed to address these issues? Lastly, we introduce a quantitative model to assess the \textit{light footprint} of individual and collective light sources on a given area. We demonstrate that integrating such a quantitative framework into LightViz can help support precise LPM and targeted mitigation strategies.

\vspace{-3mm}
\section{Background \& Related Work}
\vspace{-1mm}
\subsubsection{Light pollution sensing and mapping solutions} 
The Defense Meteorological Satellite Program ({DMSP})~\cite{dickinson1974defense} is one of the earliest satellite collecting visible and infrared images globally to study nighttime light pollution. At present, several satellite equipment such as NOAA's Advanced Very High Resolution Radiometer ({AVHRR})~\cite{cracknell1997advanced}, NASA's Moderate Resolution Imaging Spectroradiometer ({MODIS})~\cite{pagano1993moderate}, Visible Infrared Imaging Radiometer Suite ({VIIRS})~\cite{falchi2016new,nurbandi2016using}, regularly collect imagery from various spectral bands. VIIRS Day-Night Band (DNB) from NOAA-20 satellites has been proven to be effective in extracting urban areas using light intensity levels~\cite{li2022evaluating}. Using a $3000$\,Km wide swath and $0.5$-$0.9$\,$\mu$m wavelength spectrum, VIIRS imagery can detect light intensity levels for up to $2\times10^{-9}$ watts per cm$^2$ steradian, with a resolution of up to $14$\,bits~\cite{nurbandi2016using}. Researchers use the identical instrument gain setting for {Stellar Photometer} to observe long-term impacts of light pollution on marine animals~\cite{kamrowski2015influence}.

\begin{figure*}[t]
\centering
\includegraphics[width=0.98\textwidth]{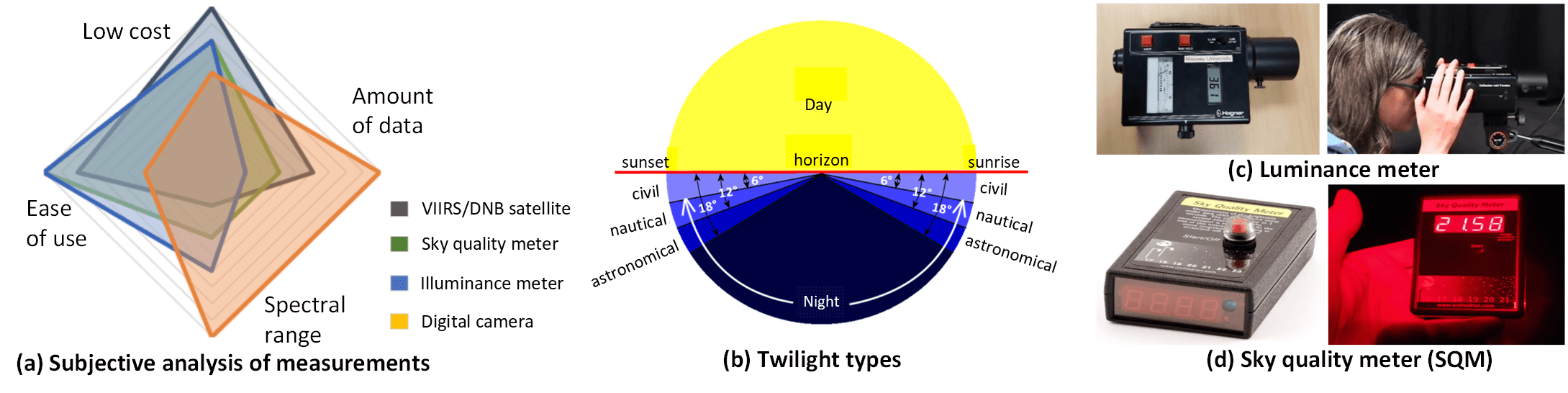}%
\vspace{-2mm}
\caption{A synopsis of existing sensors and data modalities for LPM is shown~\cite{mander2023measure,TwilightTypes}: (a) Pros and cons of various sensing modalities; (b) Different twilight types; and (c) luminance meter and SQM for single-point radiance sensing.}
\label{fig:related_work}
\vspace{-1mm}
\end{figure*}

Besides, Unmanned Ground Vehicles (UGVs)~\cite{windle2018robotic}, on-water Autonomous Surface Vehicles (ASVs)~\cite{ludvigsen2018use}, Unmanned Aerial Vehicles (UAVs)~\cite{massetti2022monitoring, tabaka2020pilot}, and air balloons~\cite{fiorentin2018minlu, walczak2021light} are used to analyze light pollution in a localized area. UAVs are also deployed to measure light emitted from a particular urban facility such as an outdoor sports complex~\cite{sielachowska2018measurement} or a park~\cite{kollath2010measuring}. In~\cite{kollath2010measuring}, drones are used to extract GPS location and a set of interesting points; human operators then measure light intensity with handheld equipment at those locations. This dynamic assessment process involves carrying heavy equipment such as DSLR cameras~\cite{kollath2010measuring, jechow2017measuring}, illuminance meters~\cite{gualuactanu2017luminance,scikezor2019light}, and spectrometers~\cite{mander2023measure, kocifaj2020emission} -- which is labor-intensive and time-consuming. For more efficient and long-term measurement, on-ground cameras and irradiance sensors are deployed at multiple locations. DSLR cameras with fisheye lenses are most commonly used to collect hemispheric photographs~\cite{windle2018robotic,hu2018association}, which are then processed by Sky Quality Meter ({SQM}) software for extracting brightness and intensity metrics pertaining to light pollution~\cite{karpinska2022device}. Fig.~\ref{fig:related_work}a provides an index of different modalities highlighting that digital cameras dominate in the spectral range, while illuminance meters are the most user-friendly devices. For nighttime, different twilight types are distinguished based on time, as shown in Fig. \ref{fig:related_work}b. This is a key factor for collecting light pollution data, as the natural background light varies at different times. Both SQM and SQM Lens Ethernet ({SQM-LE}) technologies make use of the hemispheric observation data to measure night-time brightness in magnitudes per unit area ({\tt mag/arc-sec$^2$})~\cite{mander2023measure,pun2014contributions}.
These units are qualitatively represented for visualizing spectral power distribution and wavelength sensitivity in a given region.

\subsubsection{Remote Sensing \& AI-based Monitoring Systems}
Beyond single-point sparse sensing, distributed Wireless Sensor Network (WSN) deployment is required for long-term and/or dense ecological monitoring in the field. Various {distributed WSNs} have been widely used for monitoring water quality~\cite{pule2017wireless, he2012water}, agricultural ecoparameters~\cite{pierce2008regional, rathinam2019modern}, and urban air pollution~\cite{nguyen2022wireless, liu2011developed} with promising success. Besides, an IoT-based integrated system~\cite{fang2014integrated} for regional environmental monitoring has demonstrated the effectiveness of distributed systems in environmental protection. The emission spectra of light pollution provide a unique tool for the remote diagnosis of light-polluting sources~\cite{kocifaj2020emission}. Some contemporary works show that nighttime satellite/airborne imagery and spatial vulnerability maps can be combined to find the optimal locations for deploying distributed sensor nodes for LPM or to highlight the severe locations of light pollution~\cite{lopez2023optimization,prasad2014novel,liu2021high}. Depending on the physical medium and deployment scenarios, different network topologies and communication protocols are adopted such as LoRa (Long-Range) radio waves, Bluetooth, LTEs, and WiFi. The LoRaWAN-based sensor networks are typically equipped with {TSL2591} light sensors~\cite{karpinska2022device, erwinski2023autonomous} to support long-range distributed monitoring of light pollution. However, TSL2591 sensors have a spectral range in the visible and near-infrared spectrum only and require regular human interventions for calibration. On the other hand, using SQMs in WSN nodes requires solving the inherent wavelength sensitivity and self-calibration issues, and ensuring low-power operation for high-resolution light-field mapping in real-time.  

Developing scalable and transferable light pollution monitoring networks remains a challenge; the major difficulties include: \textbf{(i)} light sensors such as luminance meters and SQMs often require periodic calibration and offline data processing, making standalone operation unfeasible; \textbf{(ii)} many light pollution zones fall into private properties where sensor node installations are challenging due to privacy issues; and \textbf{(iii)} a lack of community awareness impedes crowd-sourcing the distributed data collection process for comprehensive LPM.

\begin{figure}[h]
    \centering
    \includegraphics[width=\columnwidth]{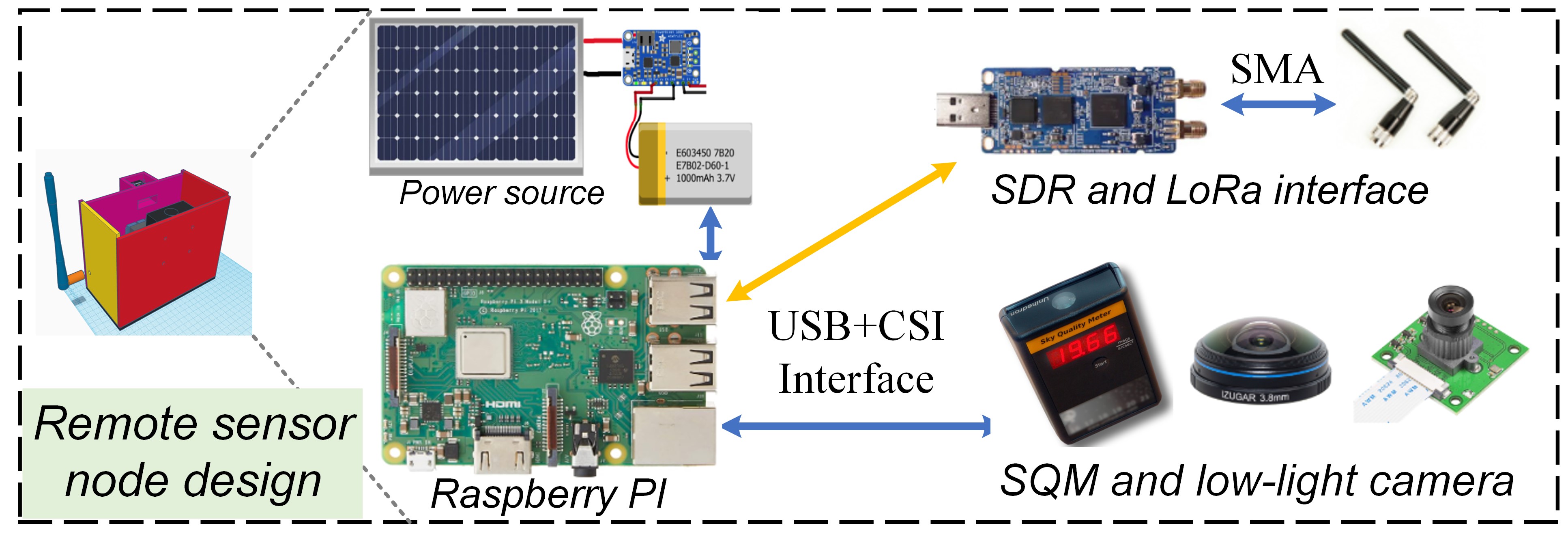}
    \includegraphics[width=\columnwidth]{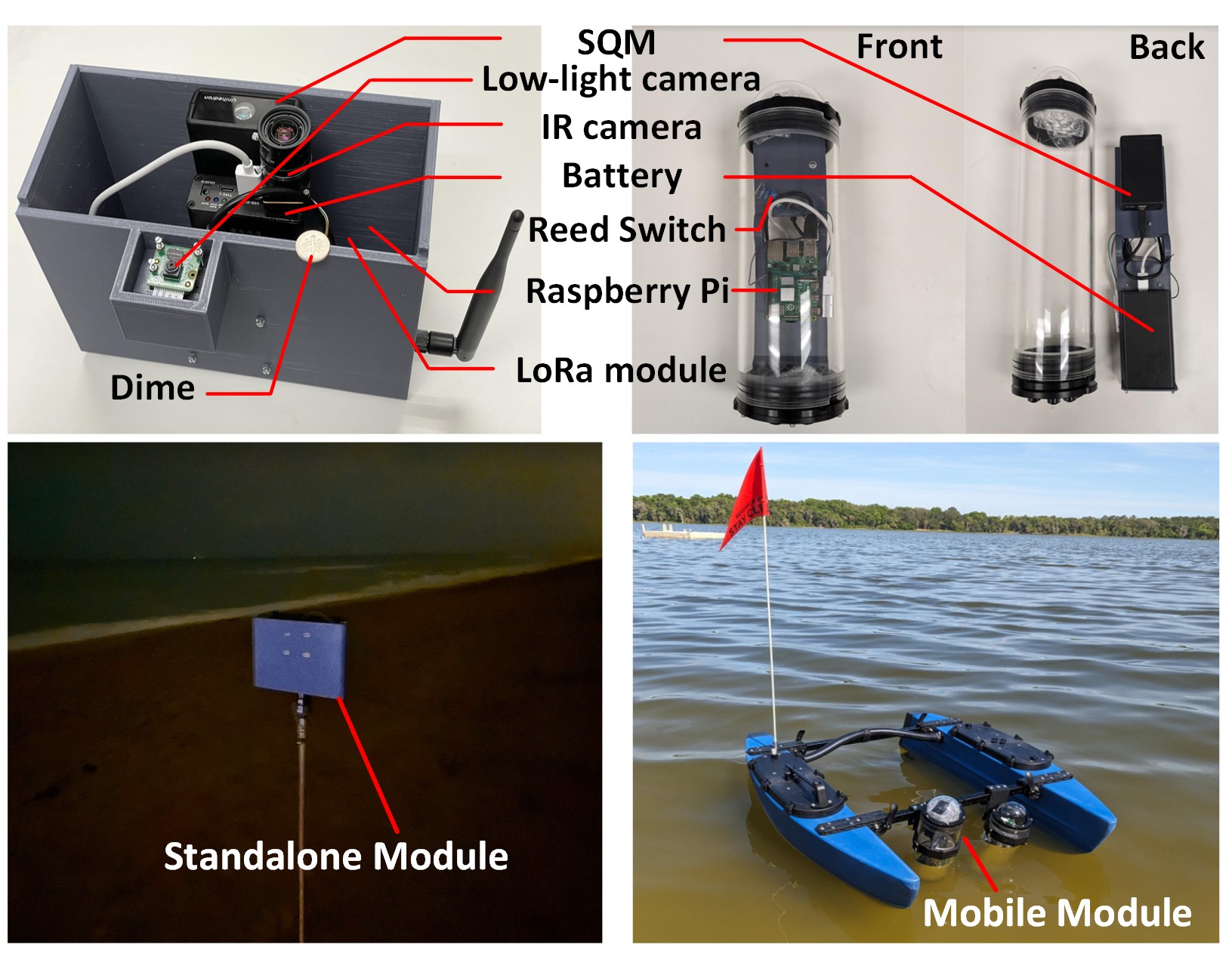}%
    \vspace{-2mm}
    \caption{The design and implementation of our remote light-field sensor node is shown. It consists of an SQM, low-light cameras, LoRa communication interface, and a portable power source. It can be used in standalone operation for overnight data collection as well in GPS-guided mobile operation on coastal waters with an ASV. 
    }%
    \vspace{-4mm}
    \label{fig:module}
 \end{figure}

\vspace{-2mm}
\section{Remote Light-field Sensing}
To overcome the issues of limited spatial resolution and manual sampling processes, we develop a compact light-field sensor module for distributed LPM~\cite{huang2024darkmeter}. As shown in Fig.~\ref{fig:module}, our remote sensor node includes an SQM, low-light photodetector and IR cameras, and LoRa communication interface -- tied to a single-board computer. It enables both single-point and mobile operation for static and dynamic light surveys respectively. Standalone module includes an SQM for light intensity measurements (in {\tt mag/arc-sec$^2$}) as well as a low-light camera and an IR camera to capture scene radiance as single-channel images. The collected data is then processed using a Raspberry PI-$4$B with $8$\,GB of RAM and subsequently transmitted to a base station via a LoRa RFM$95$ module ($915$MHz) and UHF/VHF antenna. The system is powered by a $10$\,Ah battery that ensures independent and prolonged operation, which is rechargeable with a solar panel. On the other hand, mobile robots such as ASVs are suitable platforms to integrate our sensor module for dynamic light-field sensing on water surfaces. We assemble the components of the standalone module into a watertight enclosure with a transparent dome on top to provide a window to SQM. A magnetic reed switch is added to allow the module to be powered on with a magnetic key for contactless field operation.

\begin{figure}[t]
     \centering
     \includegraphics[width=\columnwidth]{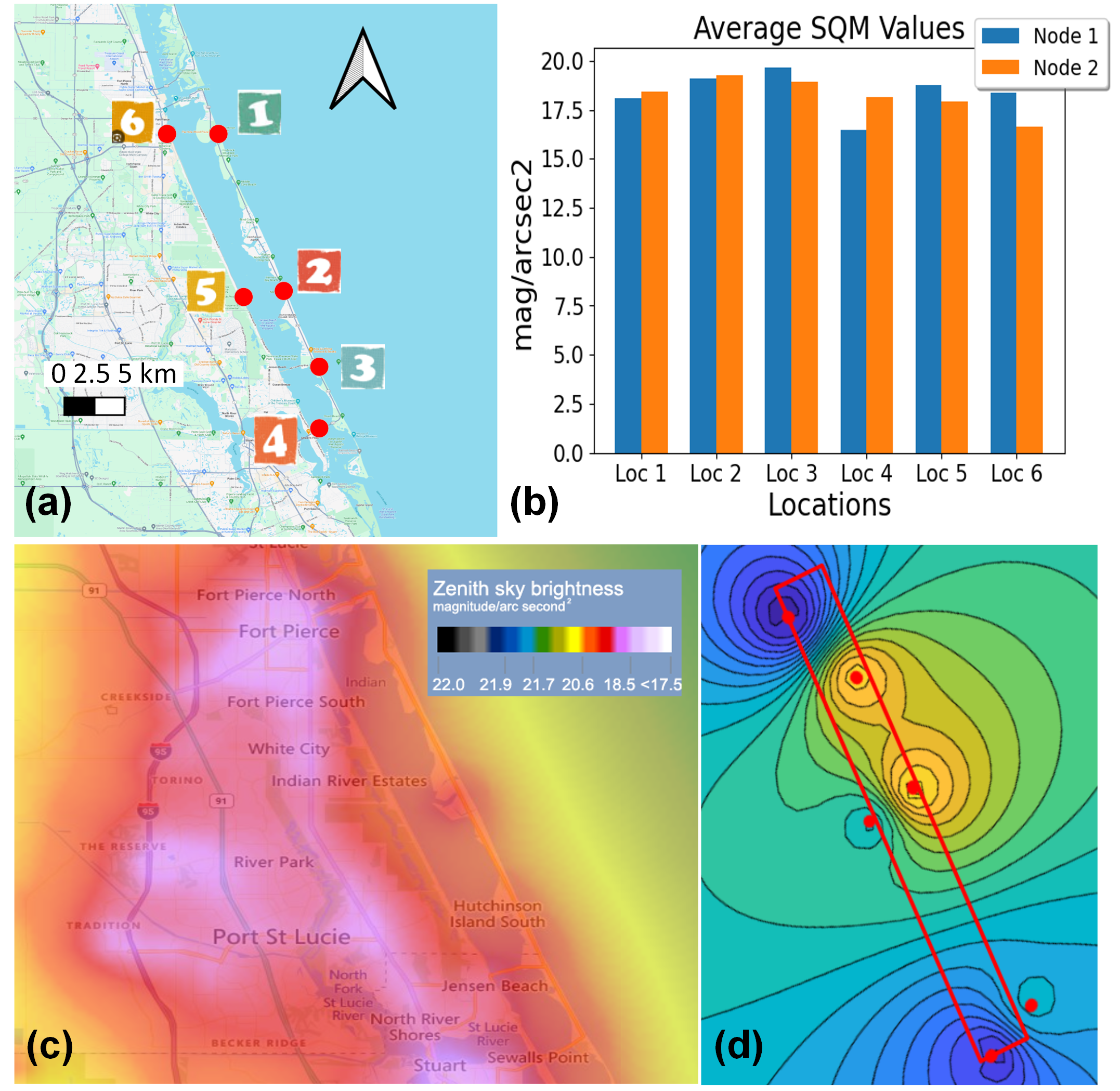}%
     \vspace{-2mm}
     \caption{Illustrations of our field deployment for distributed LPM in: (a) Jensen Beach, FL; (b) Aggregated data is shown for six specific locations; (c) Light-field map from World Atlas data; (d) Light-field contour map via spatial interpolation.}%
     \vspace{-5mm}
     \label{fig:field_exp}
 \end{figure}



Comprehensive field assessments of the remote LPM system are conducted in both beachfront communities and closed-water (lake) environments. The standalone module is first deployed at Jensen Beach, FL for long-term operation. As shown in Fig.~\ref{fig:field_exp}a, six observation locations are selected uniformly surrounding a beachfront waterbody. The exact GPS coordinates are marked and the nodes continuously capture timestamped light intensity data for $30$-minute intervals; a sample result is shown in Fig.~\ref{fig:field_exp}b. On the other hand, on-water data is collected on a BlueBoat ASV (Autonomous Surface Vehicle); GPS-guided missions are planned with multiple respective waypoints, which the ASV followed at a constant rate. The correspinding light-field maps (World Atlas data) and contour maps are shown in Fig.~\ref{fig:field_exp}c and Fig.~\ref{fig:field_exp}d, respectively. 

We compare these maps generated from the space-borne World Atlas data~\cite{falchi2016new,nurbandi2016using} with our on-ground dense light-field measurements. We investigate six SOTA interpolation methods: standard Inverse Distance Weighting (IDW) Interpolation~\cite{burrough2015principles}, Shepard Interpolation~\cite{gordon1978shepard}, Kriging Interpolation~\cite{oliver1990kriging}, Radial Basis Function (RBF) Interpolation~\cite{rippa1999algorithm}, Inverse Distance Weighting (IDW) with Variable Power~\cite{burrough2015principles}, and Nearest-Neighbor Interpolation (NNI)~\cite{rukundo2012nearest}. The goal here is to obtain continuous light-field values on these regions for comparative analyses with the World Atlas data. 

For the evaluation, we first measure the interpolation performance of SOTA methods from our field experimental data. Specifically, we select $5$ known data points along the waterbody boundary and estimate the $6$\textsuperscript{th} point (see Fig.~\ref{fig:field_exp}). The estimated values and averaged errors for a particular point ($27.438^{\circ}$, -$80.312^{\circ}$) with known SQM value is shown in Table~\ref{tab2}A. As shown, IDW and Shepard algorithms perform better than other methods for single-point measurements.  

Next, we randomly select five sample points and iteratively calculate the interpolated light-field values from our known ground measurements. As shown in Table~\ref{tab2}B, the interpolated values from all SOTA algorithms differ significantly from those of the World Atlas data (which is based on low-resolution satellite imagery~\cite{falchi2016new,nurbandi2016using}). This validates our argument that geo-spatial interpolation at such low resolution does not produce accurate light-field maps in local communities. 


\begin{table}[h]
\caption{Quantitative performance evaluation is shown for the remote sensing data from Jensen Beach, FL (see Fig.~\ref{fig:field_exp}). Six SOTA algorithms are used for interpolation: IDW~\cite{burrough2015principles}, Shepard~\cite{gordon1978shepard},  Kriging~\cite{oliver1990kriging}, RBF~\cite{rippa1999algorithm}, IDW-VP~\cite{burrough2015principles}, NNI~\cite{rukundo2012nearest}, World-Atlas~\cite{falchi2016new,nurbandi2016using}; the measurement unit is : {\tt mag/arc-sec$^2$}.}
\centering
{\small A. Estimation on an evaluation point at ($27.438^{\circ}$, $-80.312^{\circ}$).} \\
\vspace{1mm}
\scalebox{0.97}{
\begin{tabular}{l| c c c c c c}
\hline
 & IDW & Shepard & Kriging & RBF & IDW-VP & NNI  \\ 
\Xhline{2\arrayrulewidth}
Estimation & $18.36$ & $18.36$ & $18.25$ & $18.37$ & $18.36$ & $18.36$ \\
Avg. error & $1.71$ & $1.71$ & $1.71$ & $1.73$ & $1.72$ & $1.72$ \\
\hline
\end{tabular}
}

\vspace{2mm}
{\small B. Error variance ($\%$) of SOTA methods on five random samples; the World Atlas data~\cite{falchi2016new,nurbandi2016using} is used as the SQM baseline.}\\
\vspace{1mm}
\scalebox{0.92}{
\begin{tabular}{l c || c c c c c c} 
\hline
 $\#$ & W-Atlas & IDW & Shepard & Kriging & RBF & IDW-VP & NNI \\ 
\Xhline{2\arrayrulewidth}
$1$ &$19.83$ & $31.50$ & $31.50$ & $30.6$ & $37.80$ & $33.80$ &$31.50$ \\ 
$2$ &$19.28$ & $17.10$ & $17.10$ & $18.60$ & $18.67$ & $10.10$ &$17.10$  \\
$3$ &$20.31$ & $8.10$ & $8.10$ & $8.60$ & $7.10$ & $8.10$ &$8.10$ \\
$4$ &$20.72$ & $41.80$ & $41.80$ & $40.10$ & $14.60$ & $41.10$ &$41.80$ \\
$5$ &$18.92$ & $21.80$ & $21.80$ & $18.80$ & $0.10$ & $22.02$ &$21.80$ \\ \hline
$\bar{\mu}$ & $18.92$ & $21.80$ & $21.80$ & $18.80$ & $0.10$ & $22.02$ &$21.80$ \\ 
\Xhline{2\arrayrulewidth}
\end{tabular}
}
\label{tab2}
\vspace{-3mm}
\end{table}


\vspace{1mm}
\noindent
\textbf{Limitations and challenges.} While our intelligent sensing and estimation platform improves the accuracy and robustness of traditional light-field surveying~\cite{windle2018robotic,vandersteen2020quantifying}, there are some practical challenges involved for large-scale deployments. First, installing sensor nodes at scale throughout a city/state would be time-consuming and resource-intensive. The distributed light-field assessment data will remain sparse because we cannot put infinitely many light sources in any area. Secondly, it is impossible to change light source locations, types, and intensity patterns once deployed at scale. Therefore, we cannot \textit{simulate} alternate setups for finding mitigation strategies in a given community; we address these limitations in LightViz.

\section{\textbf{LightViz}: Interactive Light-field Estimation}
\begin{figure}[h]
\centering
\vspace{-4mm}
\includegraphics[width=\columnwidth]{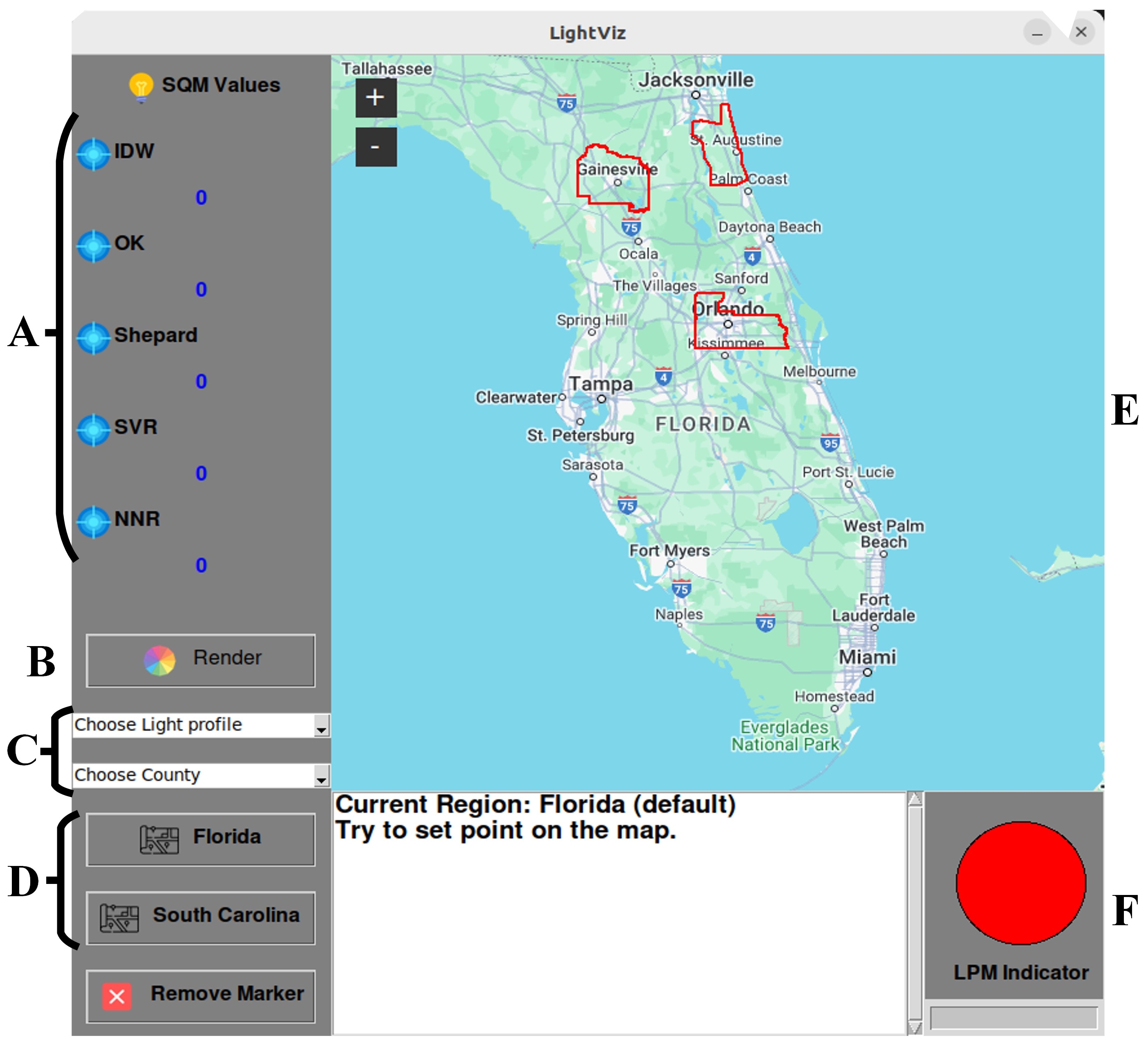}%
\vspace{-1mm}
\caption{The LightViz GUI, its parameter configurations, and functionalities are shown; (A) Spatial interpolation method selection; (B) Light-field map renderer; (C) County area and lighting profile selection; (D) State selection; (E) The geo-spatial map interface; and (F) LPM level indicator of a selected region.}
\label{fig:gui}
\end{figure}

LightViz is an interactive software interface to visualize, simulate, and map light-field for distributed LPM. The GUI (graphical user interface) of LightViz is shown in Fig.~\ref{fig:gui}; it encompasses novel features to incorporate light attenuation models, configure various light sources, select SOTA interpolation methods and map rendering techniques, and generate on-demand local and global maps for LPM. The major components of LightViz are discussed in the following sections.

\vspace{-1mm}
\subsection{Light Attenuation Model Adaptation}
\begin{figure}[t]
\centering
\includegraphics[width=0.5\textwidth]{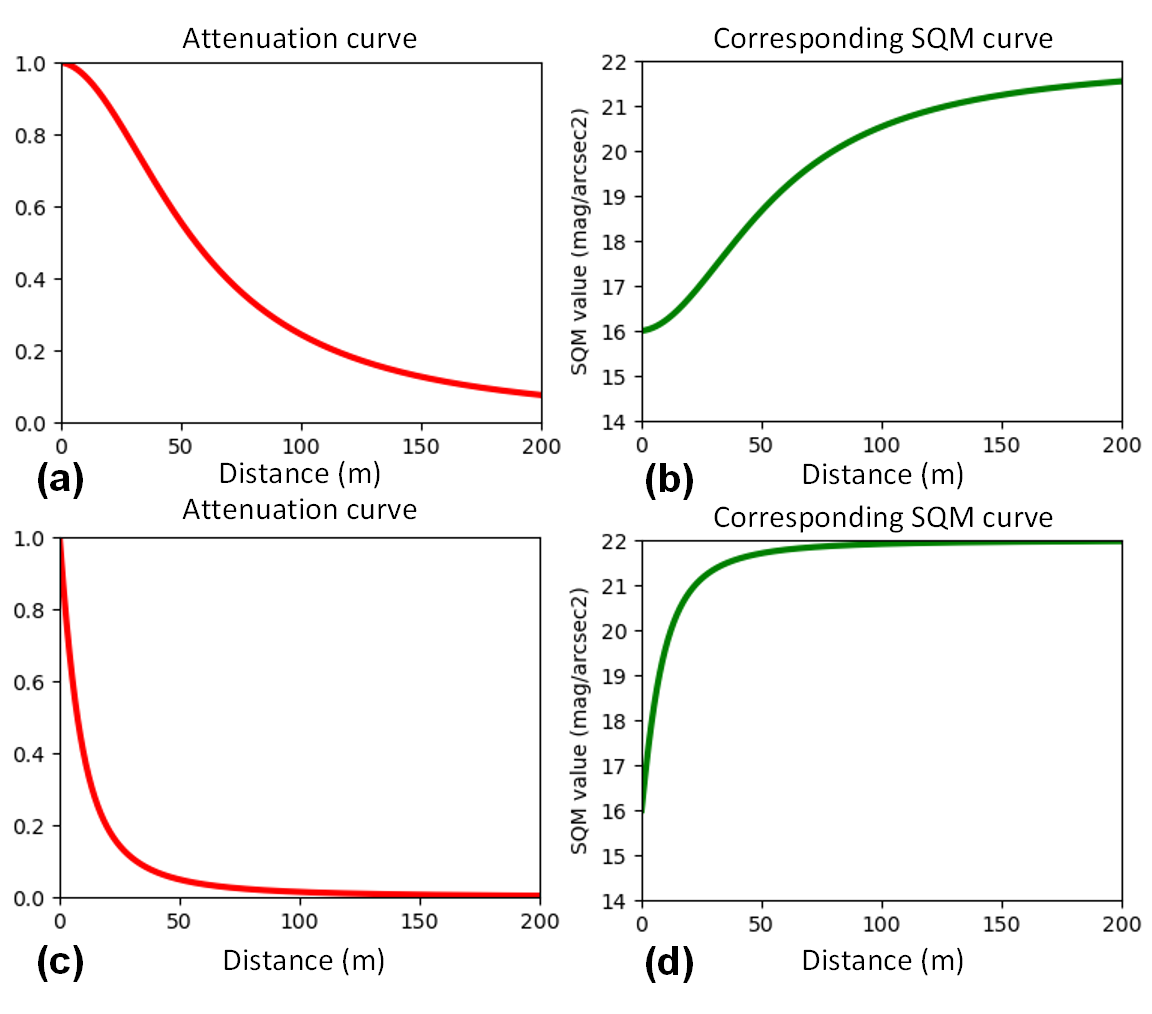}%
\vspace{-2mm}
\caption{The light attenuation curves and corresponding SQM curves are shown for (a,b) highway lights; and (c,d) rural or domestic lights.
}
\label{fig:atten}
\end{figure}

\vspace{-1mm}
The intensity of light decreases exponentially with increasing distance, while the degree of attenuation varies based on the wavelength and transmitting medium. Traditional spatial interpolation methods~\cite{mitas1999spatial} of single-point measurements lack a comprehensive light-field model to interpret this characteristic, resulting in inaccurate light-field rendering. To predict the attenuation of light radiation from a source and improve light-field rendering, we account for the attenuation of light sources to better predict light pollution. In this context, the general quadratic polynomial in the denominator allows us to represent the light attenuation function effectively~\cite{klawonn2012introduction}:
\begin{equation}
    I(d) = \frac{1}{1+c_1 \cdot \alpha d + c_2 \cdot (\alpha d)^2}\times I_0
\label{eq:light_attenuation}
\end{equation}
where $d, I, I_0, c_1, c_2$ represent the distance between the lamp and the observation point, the illumination at distance $d$, the initial illumination, and two tunable parameters, respectively. Notably, light attenuation is inversely proportional to $d^2$; the $\alpha d$ and $(\alpha d)^2$ are the first and second order smoothing terms, respectively; and $\alpha$ is the grid scaling term ($0.1$ in our case). Tuning $c_1$ and $c_2$ facilitates the fitting of attenuation curves for various light sources, allowing for precise detail and local continuity. Our tuned values of $c_1$ and $c_2$ and corresponding attenuation curves for standard U.S. highway and domestic lights are illustrated in Fig.~\ref{fig:atten}.

\vspace{-2mm}
\subsection{Light Source Type and Location Interfacing}
Streetlights are one of the primary contributors to light pollution. On different types of road surfaces, light attenuation and coverage vary significantly. For instance, on high-speed roads, such as major highways and intersections, the coverage area of light is extensive, leading to slower relative attenuation to achieve broad coverage. In contrast, on rural lanes, the coverage of streetlights is relatively limited, resulting in rapid attenuation over short distances. In LightViz, we classify the streetlamps into six major profiles based on primary U.S. road types; see Table~\ref{tab:profile}. For a selected state and county, the streetlights can be configured based on these parameter choices for light-field simulation. More importantly, stakeholders can configure custom attenuation model and light profiles seamlessly within the LightViz interface.

\begin{table}[h]
\caption{Profiles for light attenuation models based on road types for large-scale street lamps.}
\vspace{-3mm}
\begin{center}
\scalebox{0.95}{
\begin{tabular}{c|| l c c c}
\Xhline{2\arrayrulewidth}
Profile & Road Types & $I_0$ ({\tt mag/arc-sec$^2$}) & $c_1$ & $c_2$  \\  
\Xhline{2\arrayrulewidth}
$1$  & High-speed Roads
 & $16$ & $0.01$ & $0.03$ \\
$2$  & State Roads & $16$ & $0.03$ & $0.03$ \\
$3$  & County Roads & $16$ & $0.06$ & $0.03$ \\
$4$  & Municipal Roads & $16$ & $0.10$ & $0.03$ \\
$5$  & Parkways/Rural Roads & $16$ & $0.90$ & $0.60$ \\
\Xhline{2\arrayrulewidth}
\end{tabular}
}
\label{tab:profile}
\end{center}
\vspace{-5mm}
\end{table}

\vspace{-1mm}
\subsection{Interpolation Method Integration}
For light-field visualizations, we need a continuous rendering of light intensity in a given area. Since light sources are located at specific places, often with overlapping intensity regions, geospatial interpolation methods need to be incorporated to generate continuity. As shown in Fig.~\ref{fig:gui}, LightViz offers options to integrate various interpolation schemes, with IDW (inverse distance weighting)~\cite{burrough2015principles} as the default method. IDW interpolates geographic information system (GIS) data based on the distances between the interpolation point and known measurement points, using these distances as the basis for weighting. By leveraging IDW, the contributions of each light source to areas without a light source can be estimated to render a smooth and continuous light-field map.

\vspace{-2mm}
\subsection{Light-Field Map Rendering}
\vspace{-1mm}
After interpolation, a global rendering process ensures that each pixel is accurately mapped to a gradient of color representing its corresponding light-field intensity levels. This process allows users to effectively identify areas with severe light pollution and those with moderate levels. Since the unit of SQM sensors decreases with increasing light intensity levels, we reinterpret the values from $16$ (darkest) to $22$ (brightest), normalizing them from $0$ to $1$ to facilitate color mapping. The corresponding colormap spans {\tt black}, {\tt blue}, {\tt cyan}, {\tt lime}, {\tt yellow}, {\tt orange}, {\tt red}, {\tt maroon}, {\tt purple}, and {\tt white} colors -- to represent a scale of \textit{complete darkness} to \textit{maximum brightness}. Combined with a county-scale geospatial map, this allows users to visualize the extent of light pollution and track pollutant light sources in vulnerable areas of interest.

\begin{figure}[t]
\centering
\includegraphics[width=0.98\columnwidth]{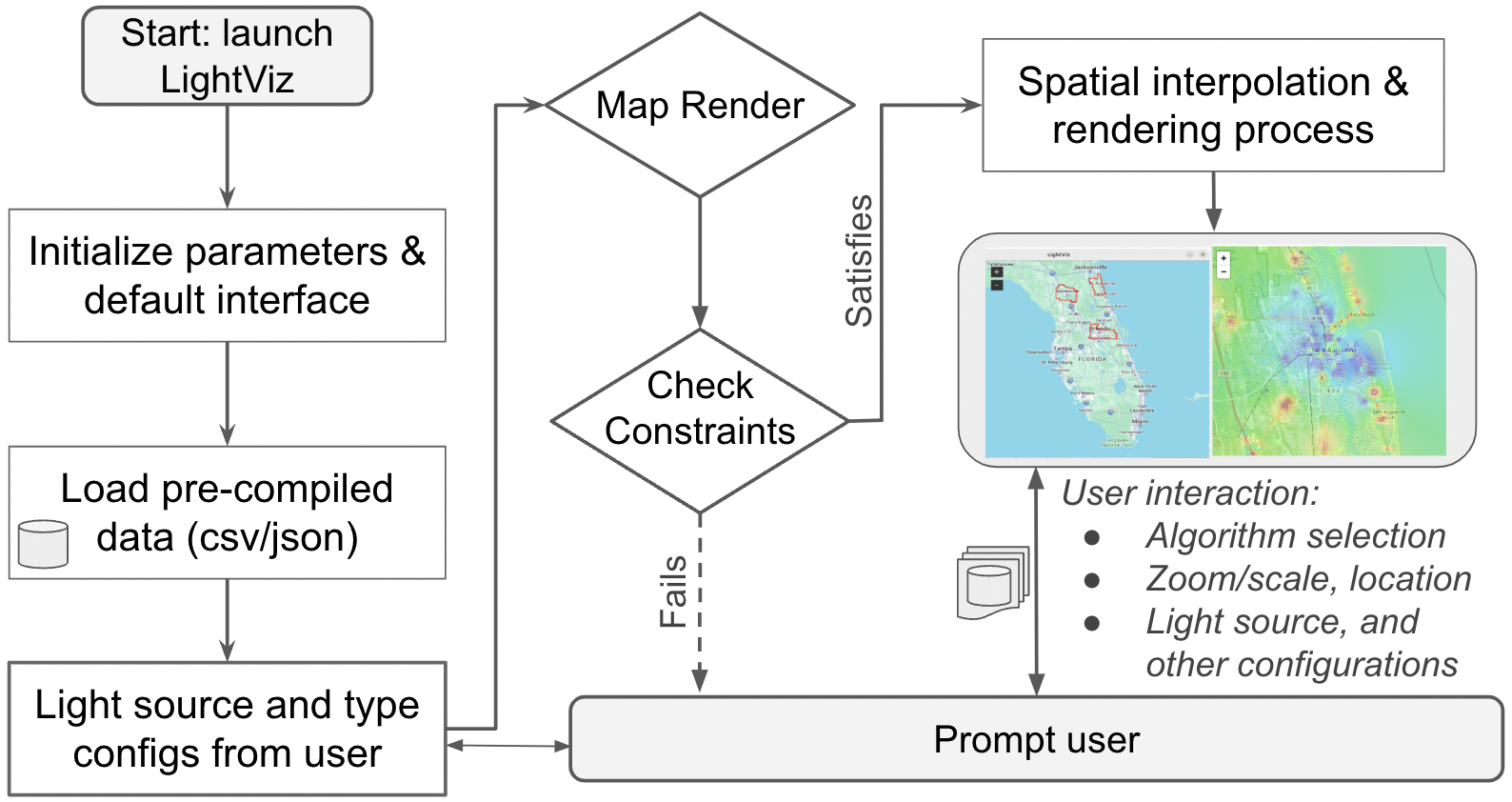}
\caption{An outline of the user commands and interaction flow for the proposed LightViz software interface is shown.}
\label{fig:flowchart}
\end{figure}

An operational outline of the LightViz software is shown in Fig.~\ref{fig:flowchart}. Once launched, users can load pre-compiled data, add new light sources and their configurations, select interpolation algorithms, and generate light-field maps in real-time. More importantly, users can interact with the map by zooming in and simulating various light-field layouts for policy identification in the selected areas.

\subsection{Attenuation configuration and measurement guidelines}
\vspace{-1mm}
Estimating accurate light attenuation curves is challenging due to the dependencies on complex physical properties of light-particle interaction in a given environment~\cite{otas2012investigation}. We follow the general practice in the literature~\cite{klawonn2012introduction} to simplify this process using curve fitting. As shown earlier in Fig.~\ref{fig:atten}, we adopt a quadratic attenuation pattern with two parameters estimated by measuring light intensities at various distances from the source using an SQM. Note that the integration of any other physically accurate model to LightViz is trivial; users can enter exact values (if known) or simulate various configurations of light source types, \eg, narrow emission spectra (LED lights), broader spectra (high-pressure sodium lights), broader spectra with rapid attenuation (incandescent bulbs), etc.

Besides, the number and arrangement of measurement points are also critical for accurate interpolation. The resolution of the interpolation depends on the number of points, which users can adjust based on their requirements. To ensure meaningful results, at least three non-collinear points are minimally required, although $20$ or more points are ideal to generate meaningful results. LightViz includes a built-in validation to check for sufficient and properly arranged points, providing user feedback when necessary (see Fig.~\ref{fig:flowchart}). This approach balances practical usability with flexibility, enabling users to model light attenuation effectively under different scenarios.

\vspace{-1mm}
\section{Experiments and Use Cases}
We demonstrate the effectiveness of LightViz with experimental analyses over a number of case studies for LPM. In the first set of experiments, we choose the St. Johns County in Florida, which is a major sea turtle conservation zone. 
In $2006$, the government initiated a two-decade conservation program; it includes an enforcement of lighting ordinance along its $41.1$ miles ($66.1$ Km) of unincorporated coastal community~\cite{lightmanagement}. The regularity of light sources, their usage and types, and available baseline data for these regions facilitate a ideal setup for our experimental evaluations.

Specifically, we develop various distributed light-field profiles for this location using LightViz and then generate the corresponding light-field maps. In the following sections, we demonstrate how to interpret those maps to monitor light pollution, identify and assess vulnerable communities, and formulate community policies for pollution mitigation.



\subsection{Light Source Layouts and Types}
In our field experimental deployments, we demonstrated how our low-power compact sensor nodes can be deployed for distributed light-field estimation and mapping. Our LightViz simulator allows us to visualize the deployed sensor data, and validate the accuracy of various geospatial interpolation algorithms. However, it is logistically challenging to conduct large-scale deployments across cities or counties for comprehensive validations. Once deployed, it also becomes challenging to alter light source types and intensity profiles to simulate various scenarios.

\begin{figure}[t]  
    \vspace{-2mm}
     \centering
     \includegraphics[width=\columnwidth]{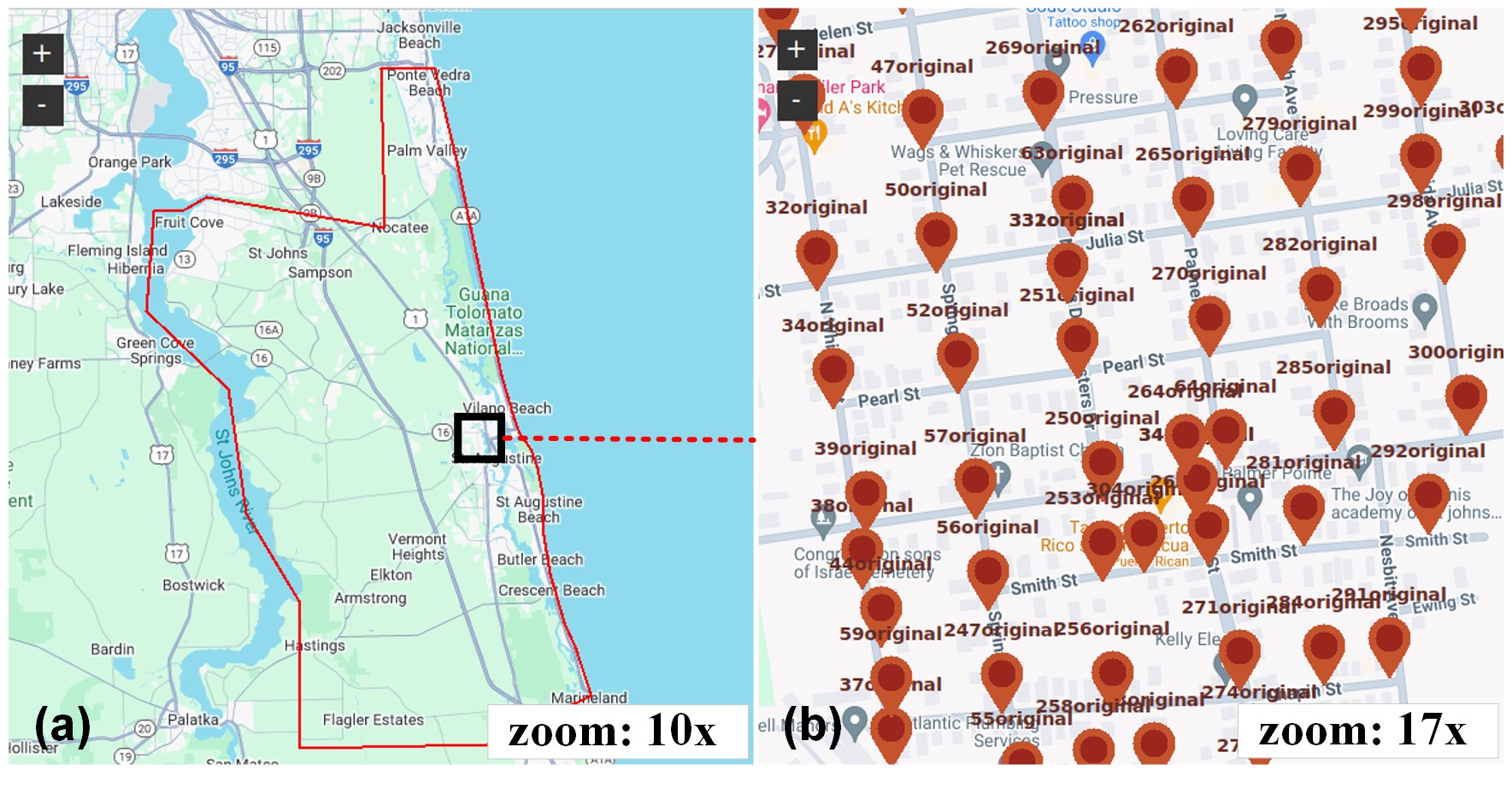}%
     \vspace{-2mm}
     \caption{A light-field profile for the St. Johns County is shown: (a) selected region in LightViz; and (b) streetlight layout configuration.}%
     \label{fig:layout}
 \end{figure}

We use the streetlight data of St. Johns County from ArcGIS Hub~\cite{StreetLightsFPL} as our initial light-source layout in LightViz. This dataset covers the downtown area as well as some sparse lighting data over the entire county as a blueprint. In LightViz, we expand this with $6,000$ additional light sources by following this blueprint for entry/exit ramps, intersections, densely populated areas, and curves with pedestrian traffic. Fig.~\ref{fig:layout} shows a snapshot of our light-source layout with all street lamps classified with light profiles presented in Table~\ref{tab:profile}. LightViz allows us to configure light sources and their profiles at the street level, extend existing layouts, and simulate various future layouts, \eg, for new residential areas.

\begin{figure}[t]  
     \centering
     \vspace{-2mm}
     \includegraphics[width=\columnwidth]{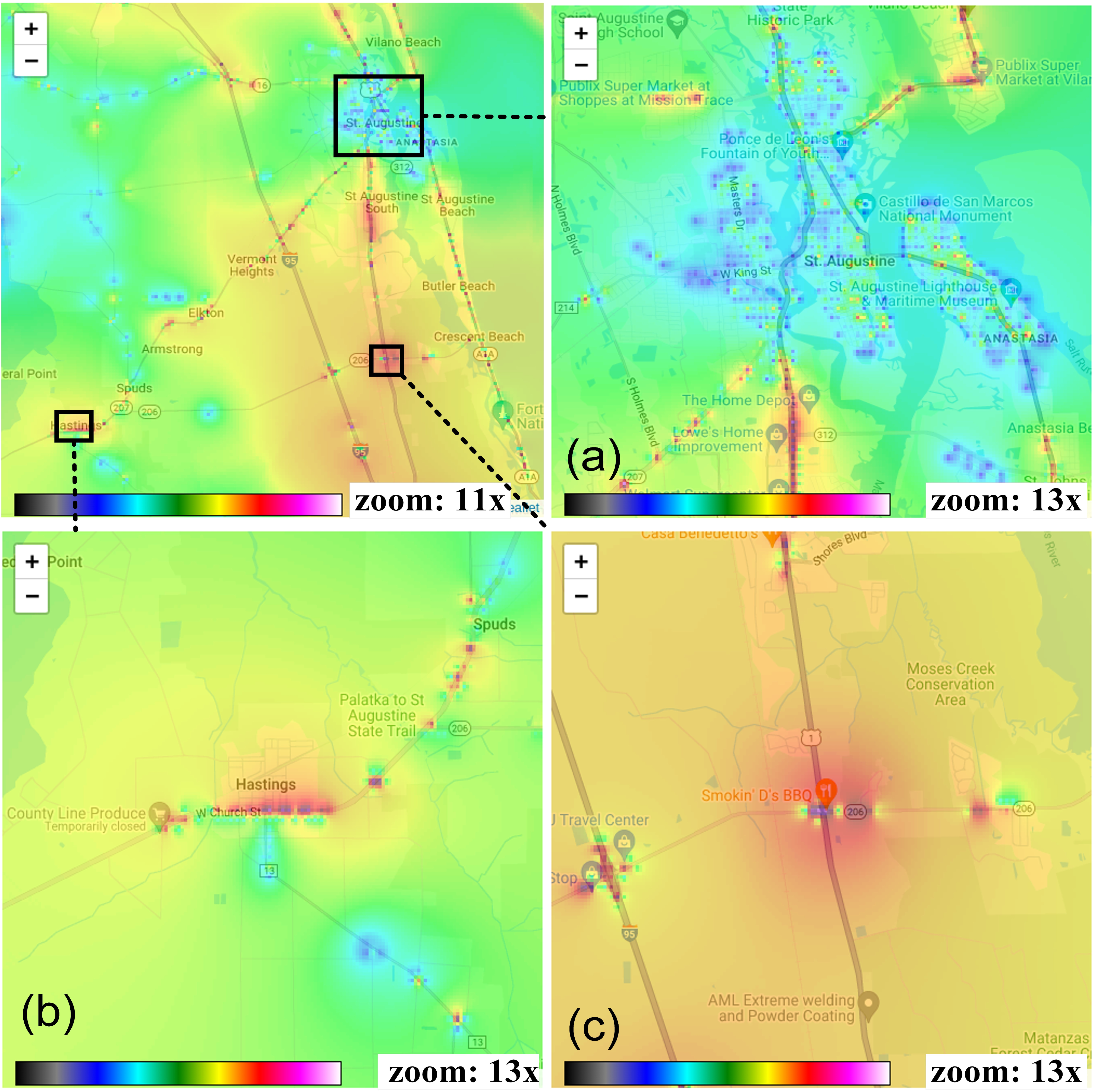}%
     \vspace{-1mm}
     \caption{LightViz rendered maps are shown for the St. Johns County; we zoom in on three specific locations: (a) the downtown area; (b) a rural area; and (c) the Moses Creek conservation Area.}%
     \vspace{-5mm}
     \label{fig:rendering_result}
 \end{figure}




\vspace{-1mm}
\subsection{Vulnerable Community Identification}
We now demonstrate that with a light source layout, we can visualize and monitor light-fields using LightViz. For the streetlight layouts discussed above, we render the geospatial light-field map of St. Johns County in Fig.~\ref{fig:rendering_result}. This map shows various regions with moderate and high levels of aggregated light intensity, which are consistent with the population density and highway routes. We highlight three particular areas in Fig.~\ref{fig:rendering_result}; first, we zoom on the downtown area where the light-field activities have a specific pattern resembling the dense city-block skeletons. The rural area and highway streetlights are also visible in Fig.~\ref{fig:rendering_result}(b) -- resembling low and high levels of light pollution, respectively.  

Moreover, light-field on the Moses Creek conservation area in  Fig.~\ref{fig:rendering_result}(c) shows vulnerable locations around the highways and major intersections. We also see the temporal pattern to be consistent with the lighting ordinance in this region. Note that these ordinances have been established through rigorous manual light-field surveys and years of conservation efforts from the local authorities. LightViz enables us to find these vulnerable communities and potential pollutant factors in minutes -- validating its utility for high-resolution LPM.

\subsection{Comparison with UAV-based aerial maps}
UAVs can effectively capture targeted aerial imagery to identify light pollution \textit{hotspot maps} from night-ground brightness (NGB) data~\cite{zhang2024evaluation,bouroussis2020assessment}. However, their utility is limited in county-scale map generation for long-term LPM due to their short flight times, geospatial image calibration with on-ground sensors, and other logistical requirements~\cite{pHodor2022detecting,fiorentin2019calibration}. Thus, community pollution tracking and effective policy identification become challenging at scale~\cite{fiorentin2019calibration,massetti2022monitoring}.

\begin{figure}[t]  
     \centering
     \includegraphics[width=\columnwidth]{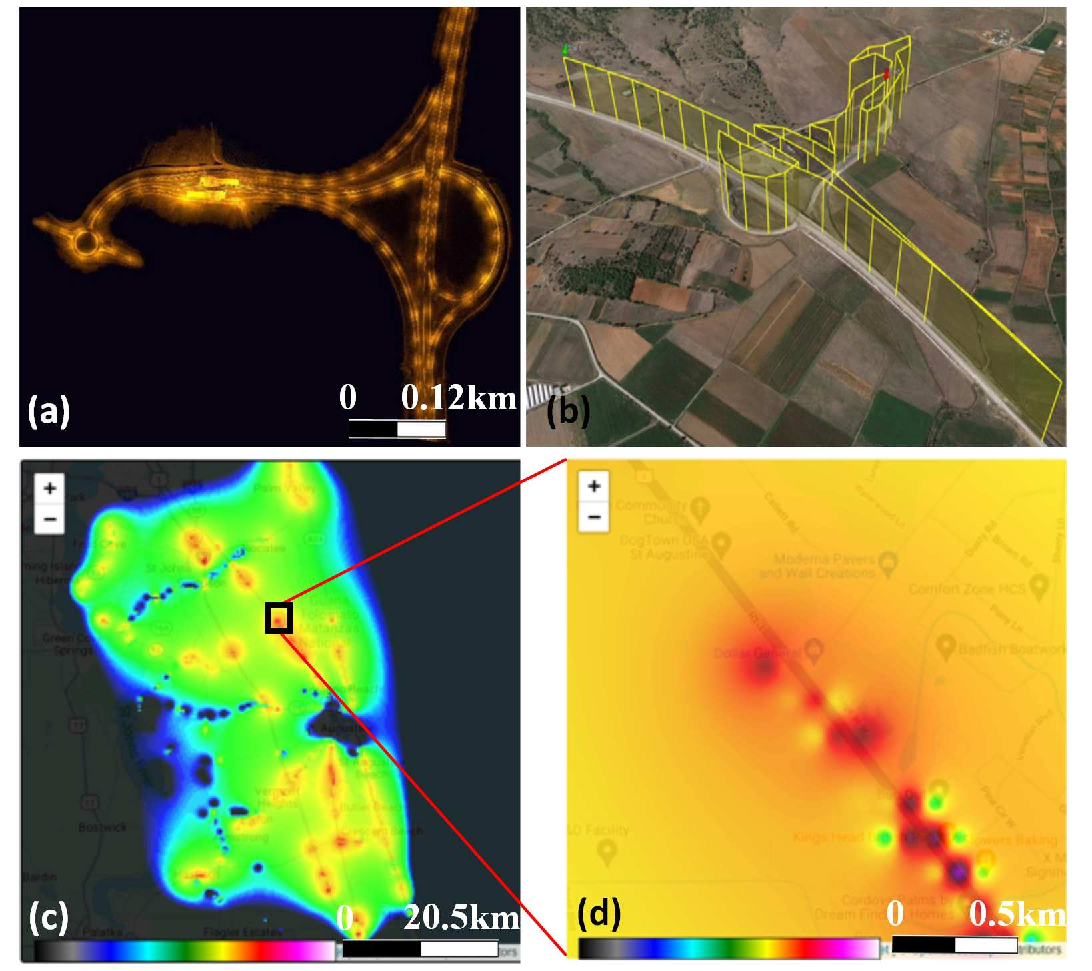}%
     \vspace{-1mm}
     \caption{{(a) Photo mosaic of an interchange from \cite{bouroussis2020assessment}; (b) the DGPM flight path over the interchange; (c) LightViz simulation of light pollution in St. Johns County; and (d) a zoom-in version illustrating seamless transitions between large-scale and small-scale views.}
     }%
     \vspace{-5mm}
     \label{fig:comparison}
 \end{figure}

In contrast, LightViz offers light source (type and placement) simulation and interactive features to generate high-resolution maps across varying scales. These capabilities facilitate the development of light pollution mitigation policies in a given community. To illustrate the differences, we compare LightViz with drone-gonio-photometer (DGPM)~\cite{bouroussis2020assessment}, an integrated technology that utilizes drones to plan trajectories and collect light pollution data. As shown in Fig.~\ref{fig:comparison}, LightViz offers broader coverage compared to DGPM and provides seamless transitions between large-scale and small-scale views. It also shows that although the DGPM map is high-resolution, the coverage area per pixel is still high. Moreover, the aerial maps are static and are not repeatable for simulating various lighting strategies and planning, as well as policymaking in large urban areas.

\subsection{Community Policy Identification}
In addition to high-resolution light-field generation and vulnerable area detection, LightViz allows effective community policy identification for conservation~\cite{gaston2012reducing,gan2023selection,liu2023assessment}. To mitigate and contain light pollution in vulnerable communities or new residential areas, it is important to know which light sources to use and where, their optimal usage patterns, and measure/track the amount of \textit{light footprint} of those sources. To this end, we demonstrate how LightViz can be used to (\textbf{i}) formulate a light source placement strategy; (\textbf{ii}) identify optimal light source types; and (\textbf{iii}) measure and track light footprints of pollutant light sources in a given area.

\begin{figure}[t]  
     \centering
     \includegraphics[width=1\columnwidth]{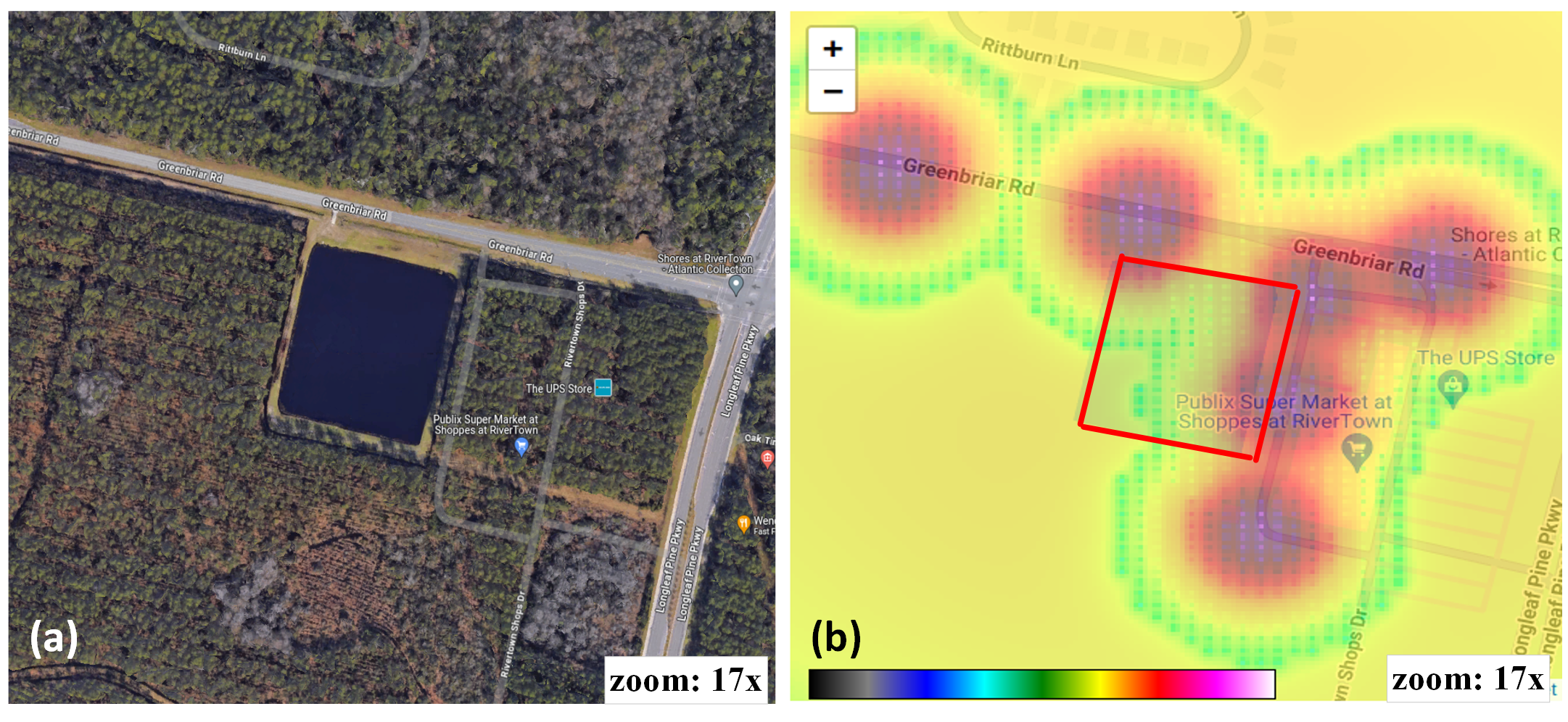}%
     \vspace{-2mm}
     \caption{(a) The satellite view of a particular lake-front area lake in St. Johns County; and (b) a light-field map featuring six streetlights surrounding the lake (marked with the red rectangle).}%
     \vspace{-5mm}
     \label{fig:original_config}
 \end{figure}

To demonstrate these features, we select a particular lake-front area in St. Johns County at ($30.055^{\circ}$, -$81.615^{\circ}$) as shown in Fig.~\ref{fig:original_config}. The nighttime illumination in this area is contributed by $6$ streetlights placements initially suggested in the following coordinates: ($30.056^{\circ}$, -$81.617^{\circ}$), ($30.055^{\circ}$, -$81.615^{\circ}$), ($30.055^{\circ}$, -$81.613^{\circ}$), ($30.053^{\circ}$, -$81.614^{\circ}$), ($30.054^{\circ}$, -$81.614^{\circ}$), and ($30.055^{\circ}$, -$81.614^{\circ}$). The initial tunable parameters are set at $(c_1, c_2) = (0.0, 0.03)$ for each light source; with this setup, the rendered light-field map is visualized in Fig.~\ref{fig:original_config}b. As seen in Fig.~\ref{fig:original_config}b, such a placement of light sources causes brightly lit areas around the lake, which will likely cause significant light pollution in the long-term.

Lets consider the scenario where we want to find a better strategy for light source placement (\ie, locations to place the light source) as well as estimate the optimal light source types to reduce the overall light-field intensity around the lake. Another objective is to quantify the amount of light in this region for long-term LPM. The constraint here is that we want to mitigate light pollution surrounding the lake (marked in red in Fig.~\ref{fig:original_config}b) while preserving a sufficient amount of lighting on the streets for drivers and pedestrians. 

To formalize this as an optimization framework, we define the free variable $\mathbf x = \{p_i,(c_{i1},c_{i2})\}$, where $p_i$ is the location of the $i^{th}$ light source, and ($c_{i1}$,$c_{i2}$) are the corresponding attenuation parameters. With $p_t$ as a location of interest, we define the objective function to model light illuminance as:
\vspace{-1mm}
\begin{equation}
f(\mathbf x, p_t) = \frac{1}{\beta} \sum_{i=1}^{N} w_{i}\times L_{i}, \quad \beta=\sum_{i=1}^{N} w_{i}. 
\vspace{-2mm}
\label{objective_fun}
\end{equation}
Here, the tunable weighting parameters $w_i$'s are defined based on the distance between $p_i$ and $p_t$ as: $w_i = 1/d(p_i,p_t)^2$. Next, we recall the intensity model (Eq.~\ref{eq:light_attenuation}) of each light source as 
\vspace{-1mm}
\begin{equation}
L_i = \frac{I_{i0}}{(1+c_{i1}\cdot y_i  + c_{i2}\cdot y_i^2)}, \quad y_i = \alpha \cdot d(p_{i}, p_{t}).
\label{eq:Li}
\vspace{-1mm}
\end{equation}

Now, to mitigate the amount of light in a point of interest $p_{t}$, we formulate the following optimization problem.
\vspace{-2mm}
\begin{align*}
\mathbf x^* &= \argmin_{\mathbf x}\, f(\mathbf x, p_t), \quad \text{\textit{subject to }} 
\end{align*}%

\vspace{-8mm}
\begin{align}
g_i(p_i)&\leq 0, \quad h_i(c_{i1}, c_{i2})\leq 0,  \quad i=1,\dots,N.
\label{eq:opt_final}
\end{align}
Here, $g_i(\cdot)$ and $h_i(\cdot)$ are the feasibility constraints for the $i^{th}$ light source placement and its parameter type, respectively. This constrained optimization problem of light source placement and type selection can be solved by standard deterministic or stochastic solvers~\cite{parsopoulos2002particle, hu2002solving}.

\subsubsection{\textbf{Light Source Placement Strategy}}
\begin{figure}[t]  
     \centering
     \includegraphics[width=1\columnwidth]{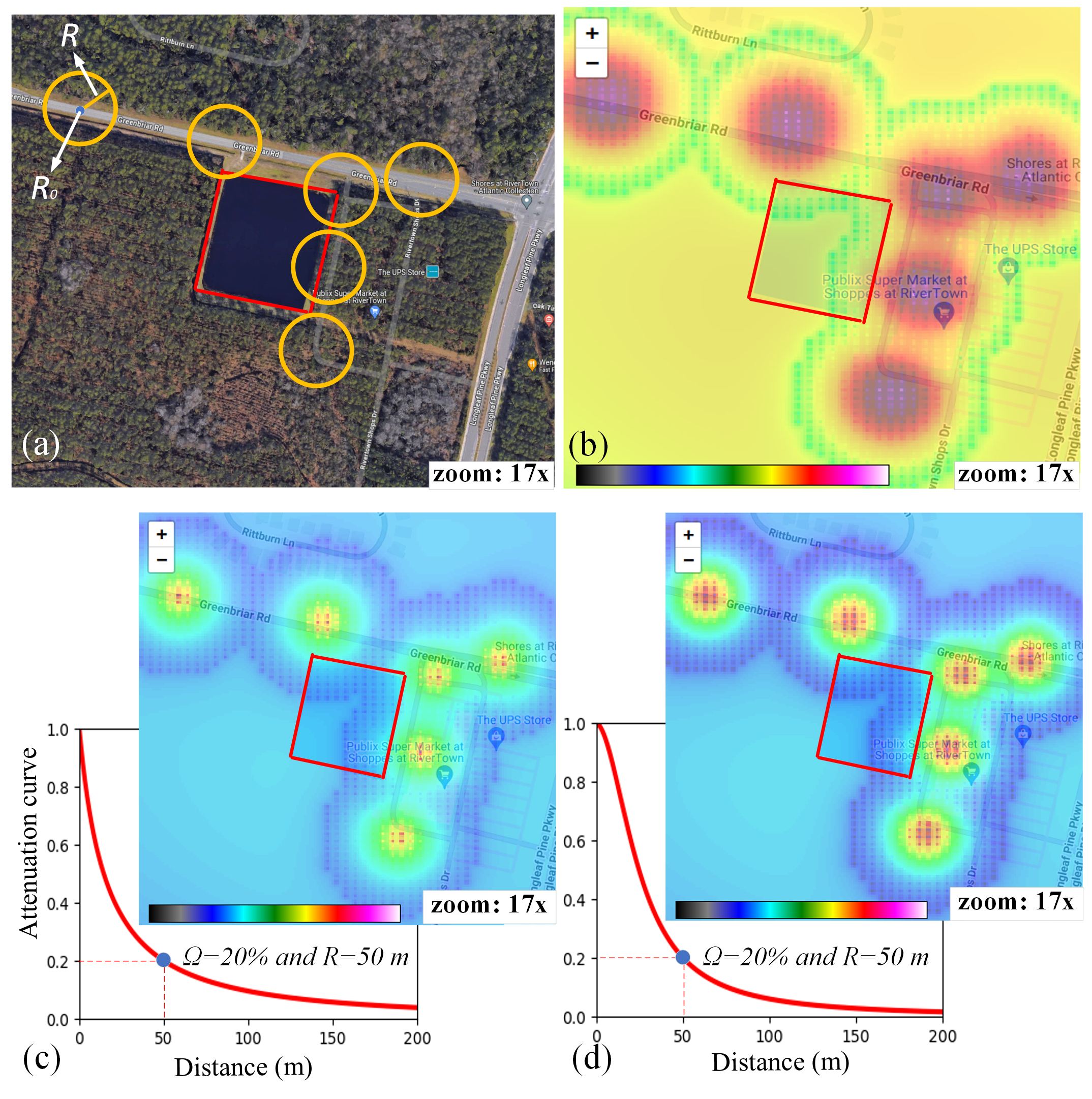}%
     \vspace{-2mm}
     \caption{(a) The $R_0$ and $R$ parameters for the six initial light source placement are annotated on the testbed area; (b) the light-field map for the optimal solution; (c,d) the light-field map for light source type identification, which tunes $c_1$ ($c_2$ fixed) and $c_2$ ($c_1$ fixed), respectively.}%
     \vspace{-2mm}
     \label{fig:PI_sol}
 \end{figure}

We now apply our optimization framework on the case shown in Fig.~\ref{fig:original_config} by parameterizing $g_i(\cdot)$ for the light source placement strategy. Specifically, we define $g_i(p_i)$ as:
\vspace{-2mm}
\begin{align}
g_i(p_i) = d(p_i,R_0) - R^2 \leq 0, \nonumber
\vspace{-2mm}
\end{align}
where $R_0$ is the initially suggested locations for each light source and $R$ is the maximum slack distance available (see Fig.~\ref{fig:PI_sol}a). That is, the light sources need to be within $R$ distance away from $R_0$ to ensure sufficient lighting on the streets. With $R=50$ meters, we solved the optimization problem by using standard {\tt SciPy API} with SLSQP method~\cite{bonnans2006numerical}.
The optimal solution $\mathbf x^*$ generates new placements for the light sources, which for our case are: ($30.0559^{\circ}$, -$81.6168^{\circ}$), ($30.0557^{\circ}$, -$81.6148^{\circ}$), ($30.0552^{\circ}$, -$81.6124^{\circ}$), ($30.0531^{\circ}$, -$81.6138^{\circ}$), ($30.0541^{\circ}$, -$81.6135^{\circ}$), and ($30.0550^{\circ}$, -$81.6133^{\circ}$). 

Based on this placement strategy, we remap the light-field as visualized in Fig.~\ref{fig:PI_sol}b -- which demonstrates considerable reduction in light intensity within the red-zone protected area, compared to the initial light-field map shown in Fig.~\ref{fig:original_config}b.

\subsubsection{\textbf{Optimal Light Source Type Identification}}
We then formulate the process of optimal light source type identification by specifying the $h_i(\cdot)$ constraint. First, we consider an attenuation model where $\Omega\%$ of the maximum brightness is maintained at a distance $R$ to ensure adequate street illumination. In this setup, we denote $h_i(c_{i1}, c_{i2})$ as:
\vspace{-1mm}
\begin{align*}
h_i(c_{i1}, c_{i2}) = -\big(\Omega\times(1+c_{i1}\cdot y_i+c_{i2}\cdot y_i^2) - 1\big) \leq 0,
\end{align*}
where $y_i = \alpha \times R$ as defined earlier in Eq.~\ref{eq:Li}. Since the parameters ($c_{i1} \cdot y_i$) and ($c_{i2} \cdot y_i^2$) are not linearly independent, we consider two different configurations: one involves adjusting $c_{i1}$ while keeping $c_{i2}$ fixed at the initial value of $0.03$; the other involves adjusting $c_{i2}$ while keeping $c_{i1}$ fixed at $0.03$. With $\Omega = 20\%$ at $R=50$ meters, we solve the Eq.~\ref{eq:opt_final} with both $g_i(\cdot)$ and $h_i(\cdot)$ constraints to get the optimal solution $\mathbf x^*$. 

For this setup, our optimal $(c_{1}, c_{2})$ values for the light sources converge to: $(0.65,0.03)$ and $(0.03, 0.154)$. The corresponding light-field maps are shown in Fig.~\ref{fig:PI_sol}c and Fig.~\ref{fig:PI_sol}d, respectively. As we can observe, attenuation curves for this  $(c_{1}, c_{2})$ profile enforces $20\%$ maximum brightness at a distance of $50$ meters. Evidently, this choice results in several standard deviations lower light-field intensities around that region. Therefore, we would install/suggest street lights whose profiles are close to the optimal $(c_{1}, c_{2})$ values for this area.





\subsubsection{\textbf{Light Footprint Measurement \& Tracking}}
Similar to measuring carbon footprint~\cite{wiedmann2008definition, hu2024review}, which quantifies the amount of greenhouse gases emitted, we propose to quantify the \textit{amount} of nighttime brightness in a given area and from a particular source. Specifically, we develop a method for quantifying light pollution as an unnatural product of human activity, enabling comparisons of the severity of light pollution across different lighting configurations. A recent work from Zhao~\etal~\cite{9484768} simulates the influence of a single-point light pollution source on residents. Other contemporary works~\cite{gaston2013ecological,gualuactanu2017luminance,bara2020magnitude} assess the impacts of light pollutant sources for correlational studies. In this work, we present mathematical formulations for two types of light footprint evaluation: (\textbf{i}) on a given area; and for (\textbf{ii}) a particular light source.

\vspace{1mm}
\noindent
\textbf{Light Footprint Evaluation of a Given Area.}  Evaluating the light footprint in a particular area involves determining the distribution and intensity of light emitted over the region. The key parameters are: the initial intensity of an impacting light source ($I_0$), the angle of emission ($\theta$); and the distance from the light source ($R$). To facilitate the calculation, we assume a light source emits light uniformly in all directions, and the initial intensity $I_0$ is distributed over a spherical surface. Then, the illuminance $E$ at a distance $R$ and angle $\theta$ from the source can be specified by the inverse square law:
\begin{equation}
\mathbf E(R, \theta) = \frac{I_0 \cos \theta}{4 \pi R^2}.
\end{equation}


To evaluate the light footprint on a specific surface, we integrate the illuminance over the area of interest. Specifically, we represent the aggregated $\mathbf E$ over a given area $A$ for all possible incident angles as follows: 
\vspace{-2mm}
\begin{equation}
\mathbf E_A = \int_0^R \int_0^{2\pi} {\mathbf E(R, \theta) } \, d\theta \, dR.
\end{equation}

\begin{figure}[t]
\centering
\includegraphics[width=\columnwidth]{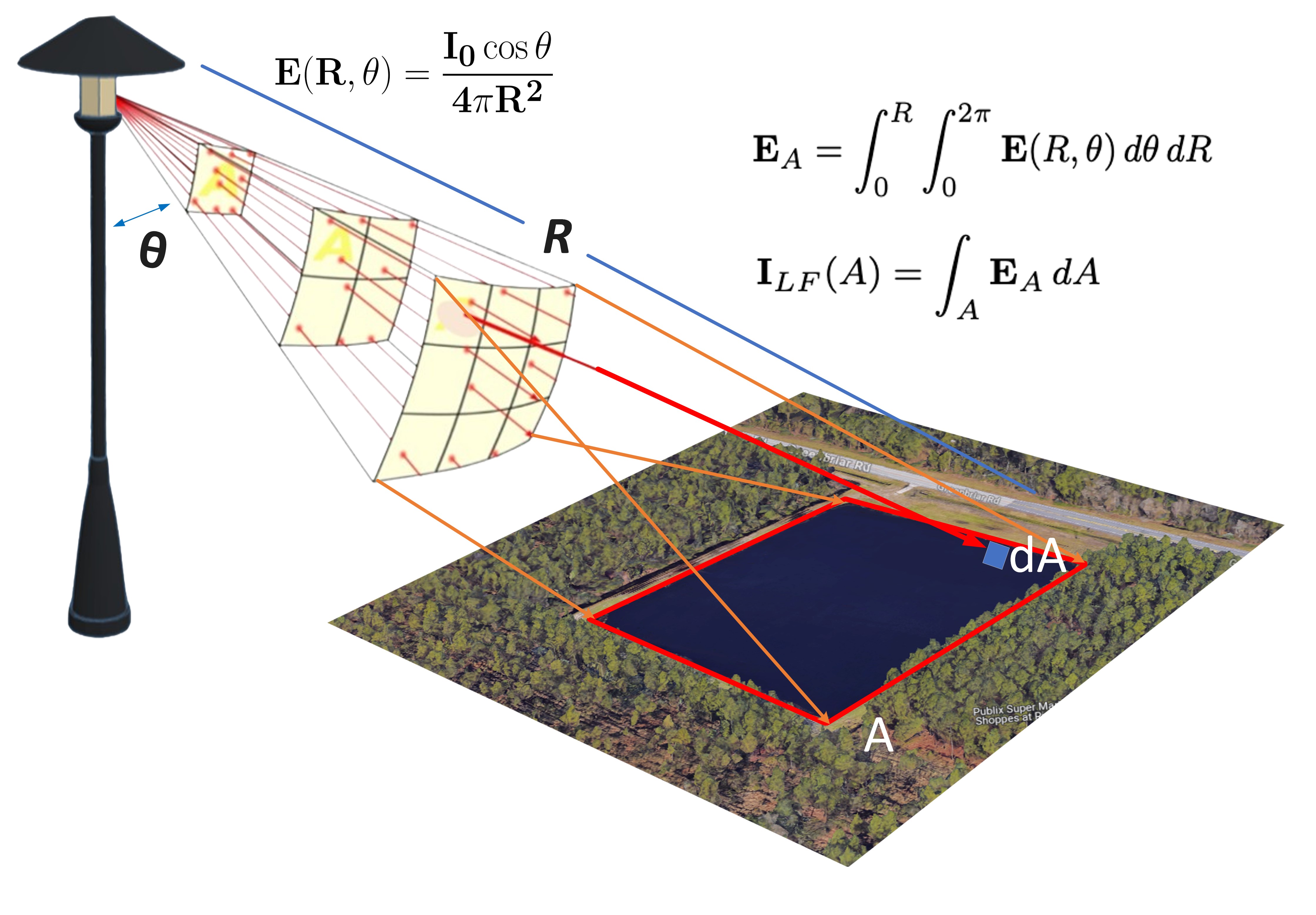}%
\vspace{-3mm}
\caption{A conceptual illustration of our light footprint evaluation for a given area: $\mathbf I_{LF} (A)$; see Eq.~\ref{eq:lf_model}. We extend this to track light footprint of a particular light source; see Eq.~\ref{eq:lf_source_model} }
\label{fig:light_footprint}
\end{figure}

Finally, we calculate the distribution of light intensity over the entire area to measure the total illuminance on the surface. As is shown in Fig.~\ref{fig:light_footprint}, we define $\mathbf I_{LF} (A)$: the total light footprint of given area $A$ as follows. 
\begin{equation}
    \mathbf I_{LF} (A) = \int_A \mathbf E_A \, dA.
\label{eq:lf_model}
\end{equation}
To facilitate the computation, we define $dA$ as $1 \ \text{m}^2$ and normalize the range of $E_A$ to $[0, 1]$.

We apply Eq.~\ref{eq:lf_model} to evaluate light footprint for the lighting configurations shown in Fig.~\ref{fig:original_config}b, Fig.~\ref{fig:PI_sol}b, Fig.~\ref{fig:PI_sol}c, and Fig.~\ref{fig:PI_sol}d. We focus on the square-shaped lake as the area of interest, and compute $\mathbf I_{LF} (A)$ for the four different lighting configurations; the results are shown in Table~\ref{tab:footprint_area}. It is noted that the placement strategy and identification of light source types successfully reduce light pollution in the vulnerable area (Fig.~\ref{fig:original_config}a) in terms of the measured light footprint for the area. Notably, the light configuration in Fig.~\ref{fig:PI_sol}d achieves the lowest footprint values, indicating that it is the best configuration among the four (with over $5.3\times$ reduction in light footprint).

\begin{table}[t]
\caption{Light footprint of the area on four lighting configurations and light-field map shown in Fig.~\ref{fig:original_config}a.}
\vspace{-3mm}
\begin{center}
\begin{tabular}{c||c|c|c|c}
\Xhline{2\arrayrulewidth}
\textbf{Map:} & Fig.~\ref{fig:original_config}b & Fig.~\ref{fig:PI_sol}b & Fig.~\ref{fig:PI_sol}c & Fig.~\ref{fig:PI_sol}d\\  
\hline
$\mathbf I_{LF} (A)$ & $17,807.36$  & $12,115.51$ & $5,692.89$ & $3,313.97$ \\
\Xhline{2\arrayrulewidth}
\end{tabular}
\label{tab:footprint_area}
\end{center}
\vspace{-3mm}
\end{table}


\vspace{1mm}
\noindent
\textbf{Light Footprint Tracking of Individual Light Sources.}
To further understand and track the contribution of individual light sources in a specified, we extend Eq.~\ref{eq:lf_model} to calculate light footprint of a pollutant light source as follows.
\begin{equation}
    \mathbf I_{LF}^i \big(p_i,c_{i1},c_{i2}, \lambda_* \big| A \big) = \int_A \mathbf E_{i}\big(p_i,c_{i1},c_{i2}\big) \, dA.
\label{eq:lf_source_model}
\end{equation}
Here, $\mathbf E_i$ is the illuminance of the $i^{th}$ light source affecting the vulnerable area $A$; $p_i$ is the location, ($c_{i1}$,$c_{i2}$) are the lighting profile parameters in LightViz, and $\lambda_*$ denote other configurable wavelength-specific light-source parameters. In LightViz, we assume that the wavelength-specific dependencies are captured by ($c_{i1}$,$c_{i2}$); nevertheless, users can integrate other $\lambda_*$ parameters for more accurate or case-specific adaptations of $\mathbf I_{LF}^i$.

\begin{table}[h]
\caption{Light footprint of the source on four lighting configurations in the vulnerable area~\ref{fig:original_config}a.}
\vspace{-3mm}
\begin{center}
\begin{tabular}{c||r|r|r|r}
\Xhline{2\arrayrulewidth}
$\#$ & Fig.~\ref{fig:original_config}b & Fig.~\ref{fig:PI_sol}b & Fig.~\ref{fig:PI_sol}c & Fig.~\ref{fig:PI_sol}d\\  
\Xhline{2\arrayrulewidth}
1 & $877.43$  & $695.78$ & $410.68$ & $151.33$ \\
2 & $3,935.03$ & $2,606.13$ & $1,184.75$ & $729.46$ \\
3 & $1,461.32$ & $1,091.40$ & $588.85$ & $248.85$\\
4 & $2,382.48$ & $1,665.98$ & $823.48$ & $408.97$\\
5 & $5,126.08$ & $3,390.31$ & $1,476.23$ & $1,021.17$ \\
6 & $4,025.03$ & $2,665.92$ & $1,208.91$ & $754.19$ \\
\Xhline{2\arrayrulewidth}
\end{tabular}
\label{tab:footprint_source}
\end{center}
\vspace{-3mm}
\end{table}

Following the light footprint evaluation of the lake area (Table~\ref{tab:footprint_area}), we apply Eq.~\ref{eq:lf_source_model} to compute light footprints for each contributing light source. We consider the same lighting configurations from Fig.~\ref{fig:original_config}b, Fig.~\ref{fig:PI_sol}b, Fig.~\ref{fig:PI_sol}c, and Fig.~\ref{fig:PI_sol}d; the results are presented in Table~\ref{tab:footprint_source}. Notably, the first streetlight has the lowest footprint value as it is the furthest from the lake, while the fifth streetlight is the most influential light source in this lighting configuration. The optimal identification of light source types, as depicted in Fig.~\ref{fig:PI_sol}d, achieves the most environmentally friendly lighting setup -- further validating the optimal policy evaluations.

\subsection{Future works}
LightViz offers several key features that make it a powerful tool for LPM, enabling high-resolution light-field mapping and real-time data collection with geospatial visualization. It also supports customizable light attenuation models, allowing users to simulate various lighting conditions and analyze their impacts. We also highlighted a particularly valuable feature for future researchers, which is to measure and track \textit{light footprints} systematically. These capabilities make LightViz an essential tool for advancing sustainable lighting solutions and ecological conservation efforts. To this end, we highlight the following research directions that will benefit the advances in LPM technology of the next generation.
\begin{enumerate}[label={$\arabic*$)},nolistsep,leftmargin=*]
\item  \textbf{AI-driven models for light-field estimation and interpolation}: By leveraging AI and machine learning algorithms trained on simulated data generated by LightViz, robust predictive models can be formulated to identify pollution trends and emerging hotspots before they become severe. Deep learning-based pipelines can also be developed to automate the classification of pollutant sources and assess their impact on nearby communities. Furthermore, real-time anomaly detection algorithms can be implemented to monitor sudden changes in lighting conditions, helping urban planners and environmentalists respond proactively. 
    \item \textbf{Refined mathematical modeling of light footprint assessment}: Further research on LightViz will facilitate the development of a more comprehensive mathematical framework for light footprint measurement and tracking. This could involve the integration of parametric and non-parametric modeling techniques to enhance the accuracy of footprint estimations. Incorporating community-specific pollutant parameters could also improve the precision of light pollution impact predictions for more sustainable urban planning. We envision that our formulations present in this paper will serve as a good baseline, with which future research works can converge to more improved formulations of light footprint assessment and tracking.
    \item  \textbf{Policy simulation and decision support systems}: Another significant extension of LightViz will be the development of an automated policy simulation and decision support system. Such a system will allow city planners and policymakers to simulate various lighting policies, such as streetlight dimming schedules, fixture replacements, and zoning regulations, to evaluate their impact before implementation. Additionally, an AI-powered recommendation engine can be developed to suggest optimized lighting configurations tailored to specific urban environments.
\end{enumerate}
These advancements will enable LightViz to transition from a human-operated mapping tool to a dynamic forecasting system that facilitates intelligent decision-making for sustainable lighting practices. Such an automated tool, in the form of a citizen app, will help engage local communities to improve awareness and collect extensive data for more effective light pollution mitigation.

\section{Conclusion}
We present \textbf{LightViz}, an interactive software interface to survey, simulate, and visualize light pollution maps in real-time for LPM. It facilitates high-resolution light-field map rendering at scale (\ie, across cities and counties) to help identify vulnerable communities and formulate effective pollution mitigation policies. LightViz incorporates seamless integration of a configurable light attenuation model, allowing users to adjust parameters for each light source individually or apply large-scale profile settings. These features help automate the tedious and error-prone manual light surveying process in practice today. A case study in St. Johns County demonstrates that LightViz can provide a detailed and informative light-field map for accurate and scalable light pollution assessment. While existing interfaces produce low-resolution sparse light-fields, LightViz achieves high-resolution light-field maps with fine-grained local details for effective community policy formulation. We validate these features with field experiments along Florida coastlines by deploying a novel remote sensing platform for distributed light-field estimation. 

Furthermore, we introduce the concept of \textbf{light footprint} that quantifies the amount of \textit{nighttime brightness} in a given area by evaluating light-field contributions from individual light sources. We are currently working on optimizing the simulator interface by user studies; we are also investigating various data-driven frameworks to automatically pinpoint vulnerable areas and pollutant light sources on the map. These improvements will further enhance the accuracy of light pollution assessments, establishing LightViz as a critical tool for sustainable city planning and environmental conservation.

\section*{Acknowledgements}
\vspace{-1mm}
This work is supported in part by the UF research grant \#$132763$. The authors would like to acknowledge the help from Dr. Blair Witherington of Inwater Research Group Inc. and Rachel Tighe at the Archie Carr Center for Sea Turtle Research. We also thank Dr. Cathi Campbell and Dr. Cynthia Lagueux at UF Department of Biology for their help and support in showcasing our early ideas and prototypes at the 2022-23 Light Pollution Management Workshop events.

\ifCLASSOPTIONcaptionsoff
  \newpage
\fi



%

%

\bibliographystyle{IEEEtran}
\bibliography{refs.bib}

\begin{thebibliography}{10}
\providecommand{\url}[1]{#1}
\csname url@samestyle\endcsname
\providecommand{\newblock}{\relax}
\providecommand{\bibinfo}[2]{#2}
\providecommand{\BIBentrySTDinterwordspacing}{\spaceskip=0pt\relax}
\providecommand{\BIBentryALTinterwordstretchfactor}{4}
\providecommand{\BIBentryALTinterwordspacing}{\spaceskip=\fontdimen2\font plus
\BIBentryALTinterwordstretchfactor\fontdimen3\font minus \fontdimen4\font\relax}
\providecommand{\BIBforeignlanguage}[2]{{%
\expandafter\ifx\csname l@#1\endcsname\relax
\typeout{** WARNING: IEEEtran.bst: No hyphenation pattern has been}%
\typeout{** loaded for the language `#1'. Using the pattern for}%
\typeout{** the default language instead.}%
\else
\language=\csname l@#1\endcsname
\fi
#2}}
\providecommand{\BIBdecl}{\relax}
\BIBdecl

\bibitem{rodrigo2023light}
J.~Rodrigo-Comino, S.~Seeling, M.~K. Seeger, and J.~B. Ries, ``Light pollution: A review of the scientific literature,'' \emph{The Anthropocene Review}, vol.~10, no.~2, pp. 367--392, 2023.

\bibitem{BENFIELD201867}
J.~A. Benfield, R.~J. Nutt, B.~D. Taff, Z.~D. Miller, H.~Costigan, and P.~Newman, ``A laboratory study of the psychological impact of light pollution in national parks,'' \emph{Journal of Environmental Psychology}, vol.~57, pp. 67--72, 2018.

\bibitem{gaston2013ecological}
K.~J. Gaston, J.~Bennie, T.~W. Davies, and J.~Hopkins, ``The ecological impacts of nighttime light pollution: a mechanistic appraisal,'' \emph{Biological reviews}, vol.~88, no.~4, pp. 912--927, 2013.

\bibitem{raap2017disruptive}
T.~Raap, J.~Sun, R.~Pinxten, and M.~Eens, ``Disruptive effects of light pollution on sleep in free-living birds: season and/or light intensity-dependent?'' \emph{Behavioural processes}, vol. 144, pp. 13--19, 2017.

\bibitem{navara2007dark}
K.~J. Navara and R.~J. Nelson, ``The dark side of light at night: physiological, epidemiological, and ecological consequences,'' \emph{Journal of pineal research}, vol.~43, no.~3, pp. 215--224, 2007.

\bibitem{boyes2021light}
D.~H. Boyes, D.~M. Evans, R.~Fox, M.~S. Parsons, and M.~J. Pocock, ``Is light pollution driving moth population declines? a review of causal mechanisms across the life cycle,'' \emph{Insect Conservation and Diversity}, vol.~14, no.~2, pp. 167--187, 2021.

\bibitem{grubisic2019light}
M.~Grubisic, A.~Haim, P.~Bhusal, D.~M. Dominoni, K.~M. Gabriel, A.~Jechow, F.~Kupprat, A.~Lerner, P.~Marchant, W.~Riley, K.~Stebelova, R.~H.~A. Van~Grunsven, M.~Zeman, A.~E. Zubidat, and F.~Hölker, ``Light pollution, circadian photoreception, and melatonin in vertebrates,'' \emph{Sustainability}, vol.~11, no.~22, p. 6400, 2019.

\bibitem{owens2020light}
A.~C. Owens, P.~Cochard, J.~Durrant, B.~Farnworth, E.~K. Perkin, and B.~Seymoure, ``Light pollution is a driver of insect declines,'' \emph{Biological Conservation}, vol. 241, p. 108259, 2020.

\bibitem{lohmann2017orientation}
K.~J. Lohmann, B.~E. Witherington, C.~M. Lohmann, and M.~Salmon, ``Orientation, navigation, and natal beach homing in sea turtles,'' in \emph{The Biology of Sea Turtles, Volume I}.\hskip 1em plus 0.5em minus 0.4em\relax CRC Press, 2017, pp. 108--135.

\bibitem{truscott2017effect}
Z.~Truscott, D.~T. Booth, and C.~J. Limpus, ``The effect of on-shore light pollution on sea-turtle hatchlings commencing their off-shore swim,'' \emph{Wildlife Research}, vol.~44, no.~2, pp. 127--134, 2017.

\bibitem{valdivia2019marine}
A.~Valdivia, S.~Wolf, and K.~Suckling, ``Marine mammals and sea turtles listed under the us endangered species act are recovering,'' \emph{PloS one}, vol.~14, no.~1, p. e0210164, 2019.

\bibitem{falchi2016new}
F.~Falchi, P.~Cinzano, D.~Duriscoe, C.~C. Kyba, C.~D. Elvidge, K.~Baugh, B.~A. Portnov, N.~A. Rybnikova, and R.~Furgoni, ``The new world atlas of artificial night sky brightness,'' \emph{Science advances}, vol.~2, no.~6, p. e1600377, 2016.

\bibitem{nurbandi2016using}
W.~Nurbandi, F.~R. Yusuf, R.~Prasetya, and M.~D. Afrizal, ``Using visible infrared imaging radiometer suite (viirs) imagery to identify and analyze light pollution,'' in \emph{IOP Conference Series: Earth and Environmental Science}, vol.~47, no.~1.\hskip 1em plus 0.5em minus 0.4em\relax IOP Publishing, 2016, p. 012040.

\bibitem{karpinska2022device}
D.~Karpi{\'n}ska and M.~Kunz, ``Device for automatic measurement of light pollution of the night sky,'' \emph{Scientific Reports}, vol.~12, no.~1, p. 16476, 2022.

\bibitem{mander2023measure}
S.~Mander, F.~Alam, R.~Lovreglio, and M.~Ooi, ``How to measure light pollution-a systematic review of methods and applications,'' \emph{Sustainable Cities and Society}, p. 104465, 2023.

\bibitem{windle2018robotic}
A.~E. Windle, D.~S. Hooley, and D.~W. Johnston, ``Robotic vehicles enable high-resolution light pollution sampling of sea turtle nesting beaches,'' \emph{Frontiers in Marine Science}, vol.~5, p. 493, 2018.

\bibitem{ludvigsen2018use}
M.~Ludvigsen, J.~Berge, M.~Geoffroy, J.~Cohen, P.~De~La~Torre, S.~Nornes, H.~Singh, A.~S{\o}rensen, M.~Daase, and G.~Johnsen, ``Use of an autonomous surface vehicle reveals small-scale diel vertical migrations of zooplankton and susceptibility to light pollution under low solar irradiance. sci adv 4: eaap9887,'' 2018.

\bibitem{massetti2022monitoring}
L.~Massetti, M.~Paterni, and S.~Merlino, ``Monitoring light pollution with an unmanned aerial vehicle: A case study comparing rgb images and night ground brightness,'' \emph{Remote Sensing}, vol.~14, no.~9, p. 2052, 2022.

\bibitem{zhang2024evaluation}
B.~Zhang, M.~Liu, R.~Li, J.~Liu, L.~Feng, H.~Zhang, W.~Jiao, and L.~Lang, ``Evaluation of urban microscopic nighttime light environment based on the coupling observation of remote sensing and uav observation,'' \emph{Remote Sensing}, vol.~16, no.~17, p. 3288, 2024.

\bibitem{bouroussis2020assessment}
C.~A. Bouroussis and F.~V. Topalis, ``Assessment of outdoor lighting installations and their impact on light pollution using unmanned aircraft systems-the concept of the drone-gonio-photometer,'' \emph{Journal of Quantitative Spectroscopy and Radiative Transfer}, vol. 253, p. 107155, 2020.

\bibitem{fiorentin2019calibration}
P.~Fiorentin, C.~Bettanini, and D.~Bogoni, ``Calibration of an autonomous instrument for monitoring light pollution from drones,'' \emph{Sensors}, vol.~19, no.~23, p. 5091, 2019.

\bibitem{pHodor2022detecting}
A.~P{\H{o}}d{\"o}r and G.~Husz{\'a}r, ``Detecting light pollution with uav, a hungarian case study,'' in \emph{2022 IEEE 22nd international symposium on computational intelligence and informatics and 8th ieee international conference on recent achievements in mechatronics, automation, computer science and robotics (CINTI-MACRo)}.\hskip 1em plus 0.5em minus 0.4em\relax IEEE, 2022, pp. 000\,197--000\,202.

\bibitem{elsahragty2015assessment}
M.~Elsahragty and J.-L. Kim, ``Assessment and strategies to reduce light pollution using geographic information systems,'' \emph{Procedia engineering}, vol. 118, pp. 479--488, 2015.

\bibitem{huang2024darkmeter}
S.-E. Huang, R.~Chen, A.~Abdullah, and M.~J. Islam, ``{Dynamic Light-field Sensing for Distributed Light Pollution Monitoring},'' in \emph{RUNE Workshop at IEEE International Conference on Robotics and Automation (ICRA)}.\hskip 1em plus 0.5em minus 0.4em\relax Yokohama, Japan: IEEE, 2024.

\bibitem{chen2002light}
W.-C. Chen, J.-Y. Bouguet, M.~H. Chu, and R.~Grzeszczuk, ``Light field mapping: Efficient representation and hardware rendering of surface light fields,'' \emph{ACM Transactions on Graphics (TOG)}, vol.~21, no.~3, pp. 447--456, 2002.

\bibitem{lightmanagement}
S.~J. Conuty, ``Light management,'' \url{https://www.sjcfl.us/light-management/}, accessed: 07-05-2024.

\bibitem{dickinson1974defense}
L.~G. Dickinson, \emph{Defense meteorological satellite program (DMSP) user's guide}.\hskip 1em plus 0.5em minus 0.4em\relax AWS, 1974, vol.~74, no. 250.

\bibitem{cracknell1997advanced}
A.~P. Cracknell, \emph{Advanced very high resolution radiometer AVHRR}.\hskip 1em plus 0.5em minus 0.4em\relax Crc Press, 1997.

\bibitem{pagano1993moderate}
T.~S. Pagano and R.~M. Durham, ``Moderate resolution imaging spectroradiometer (modis),'' in \emph{Sensor Systems for the Early Earth Observing System Platforms}, vol. 1939.\hskip 1em plus 0.5em minus 0.4em\relax SPIE, 1993, pp. 2--17.

\bibitem{li2022evaluating}
Y.~Li, Z.~Song, B.~Wu, B.~Yu, Q.~Wu, Y.~Hong, S.~Liu, and J.~Wu, ``Evaluating the ability of noaa-20 monthly composite data for socioeconomic indicators estimation and urban area extraction,'' \emph{IEEE Journal of Selected Topics in Applied Earth Observations and Remote Sensing}, vol.~15, pp. 1837--1845, 2022.

\bibitem{kamrowski2015influence}
R.~L. Kamrowski, C.~Limpus, K.~Pendoley, and M.~Hamann, ``Influence of industrial light pollution on the sea-finding behaviour of flatback turtle hatchlings,'' \emph{Wildlife Research}, vol.~41, no.~5, pp. 421--434, 2015.

\bibitem{TwilightTypes}
N.~W. service, ``Twilight types,'' \url{https://www.weather.gov/lmk/twilight-types}, accessed: 11-14-2023.

\bibitem{tabaka2020pilot}
P.~Tabaka, ``Pilot measurement of illuminance in the context of light pollution performed with an unmanned aerial vehicle,'' \emph{Remote Sensing}, vol.~12, no.~13, p. 2124, 2020.

\bibitem{fiorentin2018minlu}
P.~Fiorentin, C.~Bettanini, E.~Lorenzini, A.~Aboudan, G.~Colombatti, S.~Ortolani, and A.~Bertolo, ``Minlu: An instrumental suite for monitoring light pollution from drones or airballoons,'' in \emph{2018 5th IEEE International Workshop on Metrology for AeroSpace (MetroAeroSpace)}.\hskip 1em plus 0.5em minus 0.4em\relax IEEE, 2018, pp. 274--278.

\bibitem{walczak2021light}
K.~Walczak, G.~Gyuk, J.~Garcia, and C.~Tarr, ``Light pollution mapping from a stratospheric high-altitude balloon platform,'' \emph{International Journal of Sustainable Lighting}, vol.~23, no.~1, pp. 20--32, 2021.

\bibitem{sielachowska2018measurement}
M.~Sielachowska, D.~Tyniecki, and M.~Zajkowski, ``The measurement method of light distribution emitted from sports facilities using unmanned aerial vehicles,'' in \emph{2018 VII. Lighting Conference of the Visegrad Countries (Lumen V4)}.\hskip 1em plus 0.5em minus 0.4em\relax IEEE, 2018, pp. 1--6.

\bibitem{kollath2010measuring}
Z.~Koll{\'a}th, ``Measuring and modelling light pollution at the zselic starry sky park,'' in \emph{Journal of Physics: Conference Series}, vol. 218, no.~1.\hskip 1em plus 0.5em minus 0.4em\relax IOP Publishing, 2010, p. 012001.

\bibitem{jechow2017measuring}
A.~Jechow, Z.~Koll{\'a}th, A.~Lerner, A.~H{\"a}nel, N.~Shashar, F.~H{\"o}lker, and C.~Kyba, ``Measuring light pollution with fisheye lens imagery from a moving boat, a proof of concept,'' \emph{arXiv preprint arXiv:1703.08484}, 2017.

\bibitem{gualuactanu2017luminance}
C.~D. G{\u{a}}l{\u{a}}{\c{t}}anu, ``Luminance measurements for light pollution assessment,'' in \emph{2017 International Conference On Electromechanical And Power Systems (Sielmen)}.\hskip 1em plus 0.5em minus 0.4em\relax IEEE, 2017, pp. 456--461.

\bibitem{scikezor2019light}
{Sciezor, Tomasz and others}, ``Light pollution as an environmental hazard,'' \emph{Czasopismo Techniczne}, vol. 2019, no. Volume 8, pp. 129--142, 2019.

\bibitem{kocifaj2020emission}
M.~Kocifaj, F.~Kundracik, and O.~Bil{\`y}, ``Emission spectra of light-pollution sources determined from the light-scattering spectrometry of the night sky,'' \emph{Monthly Notices of the Royal Astronomical Society}, vol. 491, no.~4, pp. 5586--5594, 2020.

\bibitem{hu2018association}
Z.~Hu, H.~Hu, and Y.~Huang, ``Association between nighttime artificial light pollution and sea turtle nest density along florida coast: A geospatial study using viirs remote sensing data,'' \emph{Environmental Pollution}, vol. 239, pp. 30--42, 2018.

\bibitem{pun2014contributions}
C.~S.~J. Pun, C.~W. So, W.~Y. Leung, and C.~F. Wong, ``Contributions of artificial lighting sources on light pollution in hong kong measured through a night sky brightness monitoring network,'' \emph{Journal of quantitative spectroscopy and radiative transfer}, vol. 139, pp. 90--108, 2014.

\bibitem{pule2017wireless}
M.~Pule, A.~Yahya, and J.~Chuma, ``Wireless sensor networks: A survey on monitoring water quality,'' \emph{Journal of applied research and technology}, vol.~15, no.~6, pp. 562--570, 2017.

\bibitem{he2012water}
D.~He and L.-X. Zhang, ``The water quality monitoring system based on wsn,'' in \emph{2012 2nd international conference on consumer electronics, communications and networks (CECNet)}.\hskip 1em plus 0.5em minus 0.4em\relax IEEE, 2012, pp. 3661--3664.

\bibitem{pierce2008regional}
F.~Pierce and T.~Elliott, ``Regional and on-farm wireless sensor networks for agricultural systems in eastern washington,'' \emph{Computers and electronics in agriculture}, vol.~61, no.~1, pp. 32--43, 2008.

\bibitem{rathinam2019modern}
D.~D.~K. Rathinam, D.~Surendran, A.~Shilpa, A.~S. Grace, and J.~Sherin, ``Modern agriculture using wireless sensor network (wsn),'' in \emph{2019 5th international conference on advanced computing \& communication Systems (ICACCS)}.\hskip 1em plus 0.5em minus 0.4em\relax IEEE, 2019, pp. 515--519.

\bibitem{nguyen2022wireless}
H.~A. Nguyen and Q.~P. Ha, ``Wireless sensor network dependable monitoring for urban air quality,'' \emph{IEEE Access}, vol.~10, pp. 40\,051--40\,062, 2022.

\bibitem{liu2011developed}
J.-H. Liu, Y.-F. Chen, T.-S. Lin, D.-W. Lai, T.-H. Wen, C.-H. Sun, J.-Y. Juang, and J.-A. Jiang, ``Developed urban air quality monitoring system based on wireless sensor networks,'' in \emph{2011 Fifth International Conference on Sensing Technology}.\hskip 1em plus 0.5em minus 0.4em\relax IEEE, 2011, pp. 549--554.

\bibitem{fang2014integrated}
S.~Fang, L.~Da~Xu, Y.~Zhu, J.~Ahati, H.~Pei, J.~Yan, and Z.~Liu, ``An integrated system for regional environmental monitoring and management based on internet of things,'' \emph{IEEE Transactions on Industrial Informatics}, vol.~10, no.~2, pp. 1596--1605, 2014.

\bibitem{lopez2023optimization}
R.~Lopez-Farias, S.~I. Valdez, J.~Paredes-Tavares, and H.~Lamphar, ``Optimization of sensor locations for a light pollution monitoring network,'' \emph{Journal of Quantitative Spectroscopy and Radiative Transfer}, p. 108584, 2023.

\bibitem{prasad2014novel}
K.~D. Prasad, S.~Murty, and T.~Chandrasekhar, ``A novel wireless light sensing device for planetary and astronomical observations,'' \emph{Advances in Space Research}, vol.~54, no.~10, pp. 2007--2016, 2014.

\bibitem{liu2021high}
C.~Liu, Q.~Tang, Y.~Xu, C.~Wang, S.~Wang, H.~Wang, W.~Li, H.~Cui, Q.~Zhang, and Q.~Li, ``High-spatial-resolution nighttime light dataset acquisition based on volunteered passenger aircraft remote sensing,'' \emph{IEEE Transactions on Geoscience and Remote Sensing}, vol.~60, pp. 1--17, 2021.

\bibitem{erwinski2023autonomous}
K.~Erwinski, D.~Karpinska, M.~Kunz, M.~Paprocki, and J.~Czokow, ``An autonomous city-wide light pollution measurement network system using lora wireless communication,'' \emph{Sensors}, vol.~23, no.~11, p. 5084, 2023.

\bibitem{burrough2015principles}
P.~A. Burrough, R.~A. McDonnell, and C.~D. Lloyd, \emph{Principles of geographical information systems}.\hskip 1em plus 0.5em minus 0.4em\relax Oxford University Press, USA, 2015.

\bibitem{gordon1978shepard}
W.~J. Gordon and J.~A. Wixom, ``Shepard’s method of “metric interpolation” to bivariate and multivariate interpolation,'' \emph{Mathematics of computation}, vol.~32, no. 141, pp. 253--264, 1978.

\bibitem{oliver1990kriging}
M.~A. Oliver and R.~Webster, ``Kriging: a method of interpolation for geographical information systems,'' \emph{International Journal of Geographical Information System}, vol.~4, no.~3, pp. 313--332, 1990.

\bibitem{rippa1999algorithm}
S.~Rippa, ``An algorithm for selecting a good value for the parameter c in radial basis function interpolation,'' \emph{Advances in Computational Mathematics}, vol.~11, pp. 193--210, 1999.

\bibitem{rukundo2012nearest}
O.~Rukundo and H.~Cao, ``Nearest neighbor value interpolation,'' \emph{arXiv preprint arXiv:1211.1768}, 2012.

\bibitem{vandersteen2020quantifying}
J.~Vandersteen, S.~Kark, K.~Sorrell, and N.~Levin, ``Quantifying the impact of light pollution on sea turtle nesting using ground-based imagery,'' \emph{Remote Sensing}, vol.~12, no.~11, p. 1785, 2020.

\bibitem{mitas1999spatial}
L.~Mitas and H.~Mitasova, ``Spatial interpolation,'' \emph{Geographical information systems: principles, techniques, management and applications}, vol.~1, no.~2, pp. 481--492, 1999.

\bibitem{klawonn2012introduction}
F.~Klawonn, \emph{Introduction to computer graphics: using Java 2D and 3D}.\hskip 1em plus 0.5em minus 0.4em\relax Springer Science \& Business Media, 2012.

\bibitem{otas2012investigation}
K.~Otas, V.~Pakenas, A.~Vaskys, and P.~Vaskys, ``Investigation of led light attenuation in fog,'' \emph{Elektronika ir Elektrotechnika}, vol. 121, no.~5, pp. 47--52, 2012.

\bibitem{StreetLightsFPL}
J.~STAUG, ``Street lights fpl,'' \url{https://hub.arcgis.com/}, accessed: 07-05-2024.

\bibitem{gaston2012reducing}
K.~J. Gaston, T.~W. Davies, J.~Bennie, and J.~Hopkins, ``Reducing the ecological consequences of night-time light pollution: options and developments,'' \emph{Journal of Applied Ecology}, vol.~49, no.~6, pp. 1256--1266, 2012.

\bibitem{gan2023selection}
J.~Gan, S.~Guo, T.~Jin, T.~Liu, X.~Yuan, W.~Kong, R.~Qiu, and Y.~Qi, ``Selection of urban light pollution mitigation strategies based on reinforcement learning,'' in \emph{3rd International Conference on Internet of Things and Smart City (IoTSC 2023)}, vol. 12708.\hskip 1em plus 0.5em minus 0.4em\relax SPIE, 2023, pp. 800--807.

\bibitem{liu2023assessment}
B.~Liu, H.~Yang, and J.~Li, ``Assessment and mitigation of light pollution,'' \emph{Highlights in Science, Engineering and Technology}, vol.~64, pp. 132--140, 2023.

\bibitem{parsopoulos2002particle}
K.~E. Parsopoulos, M.~N. Vrahatis \emph{et~al.}, ``Particle swarm optimization method for constrained optimization problems,'' \emph{Intelligent technologies--theory and application: New trends in intelligent technologies}, vol.~76, no.~1, pp. 214--220, 2002.

\bibitem{hu2002solving}
X.~Hu, R.~Eberhart \emph{et~al.}, ``Solving constrained nonlinear optimization problems with particle swarm optimization,'' in \emph{Proceedings of the sixth world multiconference on systemics, cybernetics and informatics}, vol.~5.\hskip 1em plus 0.5em minus 0.4em\relax Citeseer, 2002, pp. 203--206.

\bibitem{bonnans2006numerical}
J.-F. Bonnans, J.~C. Gilbert, C.~Lemar{\'e}chal, and C.~A. Sagastiz{\'a}bal, \emph{Numerical optimization: theoretical and practical aspects}.\hskip 1em plus 0.5em minus 0.4em\relax Springer Science \& Business Media, 2006.

\bibitem{wiedmann2008definition}
T.~Wiedmann and J.~Minx, ``A definition of ‘carbon footprint’,'' \emph{Ecological economics research trends}, vol.~1, no. 2008, pp. 1--11, 2008.

\bibitem{hu2024review}
K.~Hu, Q.~Zhang, S.~Gong, F.~Zhang, L.~Weng, S.~Jiang, and M.~Xia, ``A review of anthropogenic ground-level carbon emissions based on satellite data,'' \emph{IEEE Journal of Selected Topics in Applied Earth Observations and Remote Sensing}, 2024.

\bibitem{9484768}
F.~Zhao, C.~Chu, R.~Liu, Z.~Peng, Q.~Du, Z.~Xie, Z.~Sun, H.~Zeng, and J.~Xia, ``Assessing light pollution using poi and luojia1-01 night-time imagery from a quantitative perspective at city scale,'' \emph{IEEE Journal of Selected Topics in Applied Earth Observations and Remote Sensing}, vol.~14, pp. 7544--7556, 2021.

\bibitem{bara2020magnitude}
S.~Bar{\'a}, M.~Aub{\'e}, J.~Barentine, and J.~Zamorano, ``Magnitude to luminance conversions and visual brightness of the night sky,'' \emph{Monthly Notices of the Royal Astronomical Society}, vol. 493, no.~2, pp. 2429--2437, 2020.

\end{thebibliography}

\vfill


\end{document}